\documentclass[a4paper,amssymb,amsmath,pra,twocolumn,floatfix,showpacs,nobalancelastpage,nofootinbib]{revtex4-1}
\usepackage{enumerate,
fancyhdr,
amsmath,
amssymb,
graphicx,
color}
\usepackage[colorlinks,allcolors=blue,bookmarks=false,hypertexnames=true]{hyperref}

\DeclareMathOperator{\sech}{sech}

\newcommand{\lllangle}[0]{\langle \! \langle \!  \langle}
\newcommand{\rrrangle}[0]{\rangle \! \rangle \!  \rangle}

\newcommand{\B}[1]{\mathbf{#1}}

\begin{document}

\title{B\"uttiker probes and the Recursive Green's Function; an efficient \\ approach to include dissipation in general configurations}

\author{\firstname{Jesse} A. \surname{Vaitkus}}
\email{jesse.vaitkus@rmit.edu.au}
\affiliation{Chemical and Quantum Physics, School of Science, RMIT University, Melbourne 3001, Australia}
\author{\firstname{Jared} H. \surname{Cole}}
\email{jared.cole@rmit.edu.au}
\affiliation{Chemical and Quantum Physics, School of Science, RMIT University, Melbourne 3001, Australia}

\begin{abstract}
An efficient and compact approach to the inclusion of dissipative effects in Non-Equilibrium Green's Function (NEGF) simulations of electronic systems is introduced. The algorithm is based on two well known methods in the literature, firstly that of the so-called Recursive Green's Function (RGF) and secondly that of B\"uttiker probes.  Numerical methods for exact evaluation of the Jacobian are presented by a direct extension to RGF which can be modularly included in any codebase that uses it presently. Then using both physical observations and numerical methods, the computation time of the B\"uttiker probe Jacobian is improved significantly. An improvement to existing phonon models within B\"uttiker probes is then demonstrated in the simulation of fully atomistic graphene nanoribbon based field effect transistors in n-i-n and p-i-n operation.
\end{abstract}


\maketitle

\section{Introduction}
The decreasing length scales of modern electronics mean that accurate descriptions of electronic transport at the mesoscopic scale are of growing interest. The semi-classical Boltzmann transport methods\cite{ALN2008} historically used to calculate system properties such as charge, spin and their respective densities and currents, break down as these systems begin to exhibit quantum behaviour. As such, a fully quantum description is required to compute these properties. For this we use the non-equilibrium Green's function (NEGF) method and the nomenclature as detailed in \cite{D2005}.

Green's function methods have worked well in describing the properties of quantum systems but the method is not without its drawbacks. When using the matrix mechanics formalism, calculation of the Green's function requires the inversion of an $N\! \times\! N$ matrix expression which scales as $\mathcal{O}(N^3)$ in time and $\mathcal{O}(N^2)$ in memory. This soon becomes computationally infeasible as the dimensions of a matrix grows, ultimately leading to a limit to the size of devices modelled. Due to this glaring problem, several algorithms have been developed to speed up these inversions.

Even though the matrix defining the Green's function is dense, the physically relevant data are typically contained close to the diagonal, and several algorithms have been developed to calculate the smallest number of elements required by a typical calculation. The Recursive Green's Function (RGF) is one such method, RGF exploits the block tridiagonal structure of the Hamiltonian by splitting the device into several `slices' which are then solved iteratively. By splitting the device up into slices the cost is reduced significantly; as an example consider a device split into $N_x$ equally sized slices, each with a corresponding block of size $N_y\! \times\! N_y$, RGF reduces the problem to one that scales as $\mathcal{O}(N_x N_y^3)$ in time and $\mathcal{O}(N_y^2)$ in memory (both a $N_x^2$ reduction in time and memory). There exist several other high performance algorithms for computing the Green's function such as FIND \cite{LD2012,LWD2013}, SPIKE\cite{PS2006}, CBR\cite{MVM+2005}, and MOR\cite{CLY+2015}, which can outperform RGF in their respective niches of parameter space but are not the focus of this paper. 

B\"uttiker probes \cite{VPG+2003,D2005} can be used to include the effects of dissipation into NEGF. The basic idea of a B\"uttiker probe is to add a fictitious `lead' to the device that can exchange energy with the system and then solve its local chemical potential such that the net current flowing through it is zero. Historically, B\"uttiker probes have been used as a phenomenological scattering model. When incorporated into the NEGF approach, such probe models are comparatively inexpensive, requiring no self-consistent iterations of the Green's functions like that of other dissipation models \cite{Lake1997,Nikonov2011}. However, B\"uttiker probes are not uniquely constrained to phenomenological devices and recent work by Greck \textit{et al.} \cite{Greck2015} explored phononic dissipation models in the context of quantum cascade lasers.

Even though these probes provide a simple way to include such effects, the canonical way to solve their chemical potentials has required the full inversion of the retarded Green's function, a clearly significant slowdown compared to the partial inversion methods commonly employed. Recently, Sadasivam \textit{et al.} \cite{Sadasivam2017} in their work on thermal transport in layered devices using phononic Green's functions provided an additional set of calculations to the recursive Green's function designed to assist in calculating the probe temperatures for B\"uttiker probes in layered devices. One limitation of the algorithm detailed by Sadisvam \textit{et al.} is that it assumes a constant chemical potential per sub block. Though suitable for some layered devices, this becomes disadvantageous when used in more diverse geometries such as those discussed by Wimmer and Richter\cite{Wimmer2009}, where the problem of optimal block tri-diagonalisation is approached. In our work we adapt the algorithm detailed by Sadasivam \textit{et al.} \cite{Sadasivam2017} for use in electronic calculations and remove the restriction on probe number and configuration, assuming only that the sparsity of the probes must be compatible with the RGF formalism. We demonstrate that such a generalisation is easily found and reduces to the result of Sadasivam et al. in the correct limits. We then demonstrate this method for several different nanoelectronic devices constructed out of armchair graphene nanoribbons using a fully atomistic description.

\section{Applying B\"uttiker probes to electronic systems}
Following an approach analogous to Ref. \cite{Sadasivam2017} for phonons, we derive an extension to the recursive Green's function to handle B\"uttiker probes for electrons. First, a brief review of Green's functions and B\"uttiker probes. The retarded electronic Green's function is given by:
\begin{equation}
G(E)^{-1} = (E+i0^+) S - H - \Sigma_\text{Leads} - \Sigma_{BP}  \label{eq:Greenfun}
\end{equation}
where $E$ is energy, $0^+$ is a small positive number to provide numerical stability, $S$ is the overlap matrix, $H$ is the Hamiltonian, $\Sigma_\text{Leads}$ is the self-energy related to the leads and $\Sigma_\text{BP}$ is an additional term added to include scattering effects. This self-energy is comprised of multiple parts that are typically spatially dependent. As an example, for phenomenological B\"uttiker probes, the retarded self energy for the $p^{\text{th}}$ probe (superscript $p$) is given by:
\begin{equation}
\Sigma_{\text{BP}}^{p}(E) = -\frac{i \hbar}{2 \tau_{\text{tot}}^p} \delta(\B{r}-\B{r}_p),
\end{equation}
where $\tau_\text{tot}^p$ is the scattering time for the $p^\text{th}$ probe located at position $\mathbf{r}_p$. Typically, in a phenomenological treatment $\tau_\text{tot}^p$ is set to some constant. In contrast, in our approach we take the scattering time to be implicitly a function of energy:
\begin{equation}
\frac{1}{\tau_\text{tot}^p(E)} = \sum\limits_{s}^{N_s} \frac{1}{\tau_s^p(E)}
\end{equation}
where $s$ counts over the $N_s$ scattering processes at site $p$. One then calculates the full self-energy as the sum over the $N_p$ probes:
\begin{equation}
\Sigma_{\text{BP}}(E) = \sum_p^{N_p} \Sigma_{\text{BP}}^{p}(E),
\end{equation}
and similarly the in-scattering matrix of these probes are given to be
\begin{equation}
\Sigma_{\text{BP}}^\text{in}(E) = \sum_p^{N_p} \Sigma_{\text{BP}}^{\text{in},p}(E) = \sum_p^{N_p} f_p \Gamma_{\text{BP}}^p(E),
\end{equation}
where $\Gamma = i(\Sigma-\Sigma^\dagger)$ is the broadening function and we define the shorthand $f_j(E)$ for the Fermi-Dirac distribution function of the $j^\text{th}$ lead:
\begin{equation}
f_j(E) = f_\text{FD}(E,\mu_j) = \left[ \exp\left(\frac{E - \mu_j}{kT}\right) + 1 \right]^{-1} \label{eq:Fermi}.
\end{equation}
The requirement of each of these probes is that the total integrated current flowing through them is zero:
\begin{align}
T_{i,k} =& \text{Tr}\left[\Gamma_i G \Gamma_k G^\dagger \right] \label{eq:Trans} \\
\tilde{I}_{i}=& \sum_{k\neq i} T_{i,k} \left[f_i(E) - f_k(E) \right]  \\
 I_{i} =& -\frac{q}{\hbar} \int \tilde{I}_i(E) \mathrm{d}E \overset{!}{=} 0 
\end{align}
$T_{i,k}$ is the transmission from lead $i$ to lead $k$, $\tilde{I}_i(E)$ and $I_i$ are the energy resolved and total integrated current flowing through lead $i$ respectively. These summations have been taken implicitly over only the probe degrees of freedom and the source and drain leads are fixed at their equilibrium chemical potentials. In experiment, one can only resolve $I_i$. For purely elastic processes these probes are further constrained to have $\tilde{I}_i(E)=0$ and this is done by inclusion of energy-dependent chemical potentials $\mu_i \rightarrow \mu_i(E)$.

\section{Numerical Approach}
To use the B\"uttiker probe method, we need to solve the chemical potential of the probes such that the total current through each is zero. There exist several methods for root-finding and for our purposes we use the Newton-Raphson approach \cite{Eyert1996}. The Newton-Raphson approach approximates functional expressions to first order by truncating their multi-variate Taylor series, thereby taking a locally linear approximation. The matrix of these partial first derivatives is called the Jacobian. The use of any Newton-Raphson-based approach requires an initial Jacobian. In many cases the Jacobian can be hard to calculate exactly and a common practice is to approximate it by finite differences. For Green's functions we find that such a treatment gives unnecessary overhead, requiring at least two RGF evaluations for each probe and requires one choose a fitting parameter. 

Instead we calculate the Jacobian exactly at each step, thus requiring only one evaluation of RGF and no fitting parameters. Though we choose to calculate and update it exactly, other techniques exist such as those that calculate it solely at the initial step and update it using convergence acceleration techniques such as the Broyden method \cite{Eyert1996}. Using the Broyden method can greatly reduce the computational complexity of the problem, though due to the approximations may take extra steps to converge. For poor guesses or crude approximations to either to the initial input state or Jacobian, a Jacobian-based approach may not converge at all. Within the scope of this work we will demonstrate that calculating the full Jacobian using the recursive Green's functions method will be more than sufficient.

Element $i,j$  of the Jacobian is defined as the partial derivative:
\begin{equation}
J_{i,j} = \frac{\partial F_i}{\partial X_j}
\end{equation}
where subscript $i$ refers to the $i^{th}$ component of the functional $F$, which for our purposes is the current through the $i^{th}$ probe and $X_j$ is the controlling parameter, which is the chemical potential of the $j^{th}$ lead. Given an initial starting guess of $X^{0}$, one computes the Jacobian at that point and then performs the step
\begin{equation}
X^{1} = X^{0} - \left( J^{0} \right) ^{-1} F^{0}
\end{equation} 
which ideally gives the root of the function. However as it is typically expected that the function is nonlinear, one must repeat the process iteratively
\begin{equation}
X^{k+1} = X^{k} - \left( J^{k} \right) ^{-1} F^{k} \label{eq:NRstep}
\end{equation}
which continues until some objective function is minimized. Given a sufficiently good starting guess, the Newton-Raphson method will converge quadratically \cite{HS1998}. There exist further so-called stabilization methods that can improve convergence in systems with singular Jacobians, noisy environments, or for systems with poor initial guesses. Examples of these methods include Levenberg-Marquardt, Steihaug conjugate gradient, and dogleg \cite{L1944,M1963,S1983,FP1963} but they are not the focus of this work and we proceed solely with the Newton-Raphson method as stated in Eq.~(\ref{eq:NRstep}). 

Previous authors \cite{VPG+2003} have chosen to calculate the elements of the Jacobian by
\begin{gather}
\tilde{J}_{i,j} = \frac{\partial f_j(E)}{\partial \mu_j} \sum_{k\neq i} T_{i,k}, \quad i = j \\
\tilde{J}_{i,j} = -\frac{\partial f_j(E)}{\partial \mu_j} T_{i,j}, \quad i \neq j
\\ J_{ij} = -\frac{q}{h}  \int \tilde{J}_{ij}(E) \mathrm{d}E.
\end{gather}
However computing this using Eq. (\ref{eq:Trans}) requires the full evaluation of the Green's function matrix to calculate these transmission values. As this is numerically unreasonable, we look to solve it in a different way. 

To reduce the complexity, Sadasivam et al. modified the B\"uttiker probe method to be compatible with the Recursive Green's Function formalism for simulation of phonon Green's functions \cite{Sadasivam2017}. Even though it has been shown that RGF is not the fastest method for matrix inversion, it is considerably faster than full inversion. Additionally if one wishes to use a quasi-Newton method such as Broyden, much faster algorithms such as FIND \cite{LD2012,LWD2013} can be used in subsequent iterations for the calculation of the current.

\subsection{Calculating the initial Jacobian}
Here we re-derive the method of Sadasivam et al. adapting it so it is applicable to electrons, rather than phonons, which can be obtained by appropriate substitution \cite{Sadasivam2017}. In this work we further develop this method to handle any number of probes of arbitrary configuration assuming only that their sparsity is compatible with the RGF formalism. For the sake of brevity in the main body of the text we discuss only the particular case of block-diagonal in-scattering self-energies, whereas the expression for block-tridiagonal in-scattering matrices can be found in Appendix \ref{sec:Gn}.

Beginning with the equation for the differential current, written in the form typically used with RGF, and suppressing the dependence on energy we have
\begin{equation}
\tilde{I}_k = \text{Tr}\left( \Sigma_k^\text{in} A - \Gamma_k G^n \right)
\end{equation}
where $\Sigma^\text{in}_k$ is the $k^{th}$ in-scattering self-energy, $\Gamma_k$ is the $k^{th}$ broadening matrix, $A$ is the spectral function and $G^n$ is the electron correlation function:
\begin{gather}
\Sigma^\text{in}_k = f_k \Gamma_k, \\
\Gamma_k = i(\Sigma_k - \Sigma_k^\dagger) \\ 
A = i(G - G^\dagger), \\
 G^n = G \Sigma^\text{in} G^\dagger.
\end{gather}
If we use the identity
\begin{equation}
\partial \text{Tr}(\B{X}) = \text{Tr}(\partial \B{X})
\end{equation}
then we can calculate the derivative of the trace by computing the trace of the derivatives as
\begin{equation}
\partial_j \tilde{I}_i = \text{Tr} \left[ \partial_j \left( \Sigma_i^\text{in} A - \Gamma_i G^n \right)  \right],
\end{equation}
where we have used the shorthand $\partial_j = \partial/\partial\mu_j$. Expanding the partial derivative explicitly gives
\begin{multline}
\partial_j \tilde{I}_i = \text{Tr} \left[   (\partial_j \Sigma_i^\text{in}) A + \Sigma_i^\text{in}(\partial_j  A) \right. \\ \left.  - (\partial_j\Gamma_i) G^n - \Gamma_i (\partial_j G^n) \right].
\end{multline}
Then, noting that both the retarded Green's function and the broadening matrices do not depend on $\mu_j$, we arrive at the expression: 
\begin{equation}
\tilde{J}_{i,j} = \text{Tr}\left(A \frac{\partial \Sigma_i^{\text{in}}}{\partial \mu_j} - \Gamma_i  \frac{\partial G^n}{\partial \mu_j} \right). \label{eq:Jacobian}
\end{equation}
As these are traces of matrix products, we can use the the following relationship when calculating them,
\begin{equation}
\text{Tr} \left( X Y \right) = \sum\limits_{i,j} (X \circ Y^T)_{ij} = \sum\limits_{i,j} (X^T \circ Y)_{ij}
\end{equation}
where $\circ$ is the Hadamard or element-wise product. When the matrices are of size $N\!\times \!N$ this reduces the number of computational operations from $\mathcal{O}(N^3)$ to $\mathcal{O}(N^2)$. Then for the particular case that $\Gamma, \Sigma^\text{in}, \text{ and } \partial_j \Sigma^\text{in}$ are block diagonal, we can compute the derivative of the Jacobian element by summing over the $N_b$ sub-blocks:
\begin{align}
\tilde{J}_{i,j} =& \sum\limits_{b=1}^{N_b} \text{Tr}\left(A_{b,b} \frac{\partial \Sigma_{i;b,b}^{\text{in}}}{\partial \mu_j} - \Gamma_{i;b,b}  \frac{\partial G_{b,b}^n}{\partial \mu_j} \right)  \\
= & \sum\limits_{b=1}^{N_b} \sum\limits_{k,l} \left(A_{b,b}^T \circ \frac{\partial \Sigma_{i;b,b}^{\text{in}}}{\partial \mu_j}  - \Gamma_{i;b,b}^T \circ  \frac{\partial G_{b,b}^n}{\partial \mu_j} \right)_{k,l} \label{eq:Jdiff}
\end{align}
where $b$ is an index running over the $N_b$ diagonal sub-blocks in RGF. $N_b$ is an index to clarify that this is for arbitrary block size and number, and not the simple example of a set of $N_x$ blocks of size $N_y \times N_y$ as described in the introduction. One can easily extend Eq.~(\ref{eq:Jdiff}) for the case of self-energies that are block-tridiagonal matrices. It is also worth noting that a very large proportion of the elements in the $\Gamma, \ \Sigma^\text{in},$ matrices are zero, which results in significant speed-up when compared to full matrices. In this derivation, no assumptions about the structure of any of these elements have been made, because they are already required to have the sparsity pattern assumed for the RGF method to be applicable.
\subsection{Calculating the Green's function derivatives}
To use the blockwise algorithm requires the calculation of both $\partial_j \Sigma_i^\text{in}$ and $\partial_j G^\text{n}$.
$\partial_j \Sigma_i^\text{in}=\Gamma_i (\partial_j f_i)$ is easily obtained in the B\"uttiker probe formalism by taking the analytic derivative of Eq.~(\ref{eq:Fermi}). All that remains is to calculate the matrices $\partial_j G_{b,b}^n$ which we will now demonstrate using a modified RGF algorithm. The essence of the technique is to calculate the derivatives of the left connected matrices which will in turn give the actual matrices. The method is directly derived from RGF and we will only reintroduce concepts required for the derivatives in this discussion. We have tried to keep our nomenclature as consistent as possible with Anatram \textit{et al.}\footnote{ A quick note on their nomenclature is that the indexing for the arbitrary matrix $X_{i+1,i}^\dagger$ would be the sub-block ``$i+1,\ i$'' of $X^\dagger$, not the conjugate transpose of the sub block ``$i+1,\ i$'' of $X$.} \cite{ALN2008} but unfortunately, they use $A$ to denote $G^{-1}$ from Eq.~(\ref{eq:Greenfun}) whereas, keeping in line with the nomenclature of Datta \cite{D2005} we use $A$ to denote the spectral function. For all instances of the $A$ from the original algorithm, we use $K$ in its stead. We begin with the definition of the left-connected electron Green's function matrix $g^{nL}$,
\begin{equation}
g_{i+1,i+1}^{nLi+1} = g_{i+1,i+1}^{rLi+1} \left( \Sigma_{i+1,i+1}^{\text{in}} + \sigma_{i+1,i+1}^{\text{in}}  \right) g_{i+1,i+1}^{aLi+1}
\end{equation}
where $g^{rL}$ are the left-connected retarded matrices, $g^{aL} = \left(g^{rL} \right)^\dagger$ are the left-connected advanced matrices, $\Sigma^{\text{in}}$ are the standard in-scattering functions and $\sigma^{\text{in}}$ are terms derived from the previous left-connected green's function: $\sigma_{i+1,i+1}^{\text{in}} = K_{i+1,i} g_{i,i}^{nLi} K_{i,i+1}^\dagger$ and the function is seeded by $g^{nL1}_{11} = g^{L1}_{11} \Sigma^\text{in}_{11} g^{L1}_{11}$. Once all $g^{nL}$ have been obtained, setting the last block of the full electron Green's function matrix $G_{N,N}^n$ by $G_{N,N}^n = g_{N,N}^{nLN}$ one can obtain the full electron Green's function $G^n$ by the recursive relation:
\begin{multline}
G_{i,i}^n = g_{i,i}^{nLi} + g_{i,i}^{rLi} \left( K_{i,i+1} G_{i+1,i+1}^n K_{i+1,i}^\dagger   \right) g_{i,i}^{aLi} \\
 -\left( g_{i,i}^{nLi} K_{i,i+1}^\dagger G_{i+1,i}^\dagger + G_{i+1,i} K_{i+1,i} g_{i,i}^{nLi} \right)
\end{multline}
the process for calculating the derivative of the electron Green's function $\partial_j G^n$ is exactly the same, though with the derivative $\partial_j \Sigma^\text{in}$ instead of $\Sigma^\text{in}$, this is because $g^{rL}$ do not depend on the chemical potentials, one calculates the left-connected electron Green's function derivatives $\partial_j g^{nL}$ by
\begin{equation}
\frac{\partial g_{i+1,i+1}^{nLi+1}}{\partial \mu_j}  = g_{i+1,i+1}^{rLi+1} \left( \frac{ \partial \Sigma_{i+1,i+1}^{\text{in}} }{\partial \mu_j} + \frac{ \partial \sigma_{i+1,i+1}^{\text{in}} }{\partial \mu_j}  \right) g_{i+1,i+1}^{aLi+1} \label{eq:gnderiv}
\end{equation}
where $\partial_j \sigma_{i+1,i+1}^{in} =  K_{i+1,i} \left(\partial_j g_{i,i}^{nLi} \right) K_{i,i+1}^\dagger$ and  similarly the function is seeded by $\partial_j g^{nL1}_{11} = g^{L1}_{11} (\partial_j \Sigma^\text{in}_{11}) g^{L1}_{11}$. Setting $\partial_j G_{N,N}^n = \partial_j g_{N,N}^{nLN}$ one then obtains $\partial_j G^n$  by the recurrence relation: 
\begin{multline}
\frac{\partial G_{i,i}^n}{\partial \mu_j} = \frac{\partial g_{i,i}^{nLi}}{\partial \mu_j} + g_{i,i}^{rLi} \left( K_{i,i+1} \frac{\partial G_{i+1,i+1}^n}{\partial \mu_j} K_{i+1,i}^\dagger   \right) g_{i,i}^{a Li} \\
 -\left( \frac{\partial g_{i,i}^{nLi}}{\partial \mu_j} K_{i,i+1}^\dagger G_{i+1,i}^\dagger + G_{i+1,i} K_{i+1,i} \frac{\partial g_{i,i}^{nLi}}{\partial \mu_j} \right). \label{eq:Gnderiv}
\end{multline}
What is remarkable about this approach is that it uses all the information from the initial RGF calculation, with no new inverses needing to be calculated, only matrix products. The algorithm as presented here has assumed for the sake of brevity that the in-scattering matrices are block diagonal, as is commonly assumed in the literature. However this is not a hard restriction and the algorithm is easily extensible to handle block tridiagonal in-scattering matrices which we have included in appendix \ref{sec:Gn}, again noting that the only difference between $G^n$ and its derivative is the substitution $\Sigma^\text{in} \rightarrow \partial_j \Sigma^\text{in}$.

Equations (\ref{eq:gnderiv}, \ref{eq:Gnderiv}) can be modularly included into any framework that already uses recursive Green's functions. For layered heterostructures, explicitly setting all chemical potentials in each layer to be equal these formulae reduce to the results in Sadasivam \textit{et al.} \cite{Sadasivam2017} and due to the flexibility of the description, any other such symmetry can also be included such as mirror or rotational symmetry. 

For this work we have used the recursive Green's function algorithm because of its efficient reuse of previous data to calculate the electron Green's function derivatives. Similarly, any algorithm that can efficiently reuse the outputs of the retarded Green's function calculations to compute several electron Green's functions would also be useful to calculate the Green's function derivatives.

\subsection{Approximating the Jacobian}
In this section we will discuss an approximation to the Jacobian, taken in the high and low temperature limits. First we re-introduce the differential version of the Jacobian (\ref{eq:Jacobian}) but split into $\tilde{J}^\text{L}, \ \tilde{J}^\text{NL}$ corresponding to the local and non-local parts respectively,
\begin{equation}
\tilde{J}_{i,j}^\text{L} = \text{Tr}\left(A \frac{\partial \Sigma_i^{\text{in}}}{\partial \mu_j} \right), \quad  \tilde{J}_{i,j}^\text{NL} = - \text{Tr}\left( \Gamma_i  \frac{\partial G^n}{\partial \mu_j} \right). \label{eq:Jsplit}
\end{equation}
It can be seen that $\tilde{J}_{i,j}^\text{L}$ is diagonal by writing:
\begin{align*}
\Sigma^\text{in} &= \sum\limits_{i} \Sigma^\text{in}_i = \sum\limits_{i} f_i \Gamma_i \\
\frac{\partial \Sigma^\text{in}}{\partial \mu_j} &= \sum\limits_{i}\frac{\partial \Sigma^\text{in}_i}{\partial \mu_j} = \sum\limits_{i} \frac{\partial (f_i \Gamma_i)}{\partial \mu_j}  = \frac{\partial  f_i}{\partial \mu_i} \Gamma_i  \delta_{ij}
\end{align*}
After some scrutiny, it becomes clear that $\tilde{J}^\text{L}$ is the dominant contribution to the Jacobian and we calculate it exactly using the methods described previously. This can be observed by expanding Eq. (\ref{eq:Jsplit}) explicitly as
\begin{equation}
\tilde{J}_{i,j}^\text{L} =\delta_{ij} \text{Tr}\left(\Gamma_j A  \right) \partial_j f_j, \quad  \tilde{J}_{i,j}^\text{NL} = - \text{Tr}\left(\Gamma_i A_j \right) \partial_j f_j . \label{eq:Jsplit2}
\end{equation}
and that one can see that both $\tilde{J}_{i,j}^\text{L}$ and $\tilde{J}_{i,j}^\text{NL}$ are weighted products of a broadening matrix and a spectral function. $\tilde{J}_{i,j}^\text{NL}$ uses the partial spectral function $A_j$ whereas $\tilde{J}_{i,j}^\text{L}$ uses the full spectral function $A = \sum_j A_j$. Calculating $\tilde{J}_{i,j}^\text{L}$ and its integral are both expedient and computationally inexpensive due to the retarded Green's functions and self-energies (and hence the broadening matrices) being stored in memory, and the derivative of the Fermi-Dirac function being analytic. However, to calculate $\tilde{J}_{i,j}^\text{NL}$ requires the product of the broadening matrix with an electron Green's function derivative. Although it is inexpensive to calculate these terms individually, it becomes prohibitively expensive when the number of probes and energy points become very large. Instead we will look to approximate them by their dominant contributions.

We would like to rationalise the necessity of making an approximation to $\tilde{J}^\text{NL}$ as follows: consider that in our observations, convergence of the probe currents requires typically fewer than 10 Newton-Raphson iterations when using the exact Jacobian. To compute the current at each iteration requires the computation of $N_E$ electron Green's functions, one for each energy point. However, the calculation of $\tilde{J}^\text{NL}$ requires $N_p \times N_E$ electron Green's function derivatives as well as $N_p$ sets of products and summations (where $N_p$ is the number of probes). This means that computation of $\tilde{J}^\text{NL}$ is $N_p$ times larger than a single current evaluation, and efforts like stabilisation \cite{L1944,M1963,S1983,FP1963}, approximation of the Jacobian and/or use of quasi-Newton methods such as Broyden \cite{Eyert1996}, at the cost of further iterations become well justified. We will not focus on quasi-Newton methods as they have been well studied in the literature. Instead in an effort to reduce the computational load of the Newton-Raphson method, we look to approximate $\tilde{J}^\text{NL}$.

First we observe that the derivative of the Fermi-Dirac distribution function is given by
\begin{align}
\frac{\partial f_j(E)}{\partial \mu_j} =&  \frac{1}{4 k_B T} \sech^2\left(\frac{E-\mu_j}{2 k_B T} \right).
\end{align}
The function itself is peaked at the chemical potential of the probe and then decays inversely proportionally to temperature with approximately $50\%$ of its area being located between $\mu_j\pm k_B T$. As $k_BT$ approaches zero, the Fermi-Dirac distribution derivative approaches the Dirac delta function. At very low temperatures, energy resolution becomes a major concern as one is likely to miss sharply peaked features. As such, treating it as a delta function alleviates  concerns of the numerical accuracy and also results in a $N_E$ speed-up. 

The second limit we wish to discuss is that of high temperature, physically-motivated self-energies such as those related to acoustic phonons increase their broadening proportionally to temperature. In such a case we assume that the product of the self-energy and the partial density of states varies slowly over the energy range of interest. We note that because most self-energies can be decomposed into the sum of Lorentzians, we expect that unless near resonance, the partial density of states will decay slowly with an inverse quadratic behaviour. We use these observations as justifications to replacing the derivative of the Fermi-Dirac distribution function with a delta function and the integral reduces to
\begin{equation}
\int\limits_{-\infty}^{\infty} \text{Tr}\left[ \Gamma_i(E) \frac{\partial G^n(E)}{\partial \mu_j} \right] dE \approx \text{Tr} \left[\Gamma_i(\mu_j) A_j(\mu_j)\right]
\end{equation}
where $A_j$ is the partial spectral function $A_j = G \Gamma_j G^\dagger$ that can be computed using the recursive Green's function algorithm under the substitution $\Sigma^\text{in}\rightarrow \Gamma_j$. In intermediate regimes, or when close to resonance, we do not expect a delta limit to be a viable approximation but rather computing the (weighted) average of $\text{Tr}(\Gamma_i A_j)$ using a small window centered about the probe chemical potential plus/minus a few $k_BT$ could be used. For all calculations hereafter we operate using a fixed temperature of $300 \text{ K}, (k_B T \approx 26 \text{ meV})$ and proceed using the delta function approximation. Lastly, we note that in some cases $\partial_j G^\text{n}$ can be very small due to the specific scattering model, delta approximation or not. To prohibit the chemical potentials not running away in such a situation, we clip step sizes to a maximum of 100meV. For the cases studied our results demonstrate a clear resilience to the approximations detailed. 

\section{Phonon Dissipation Models}

The derivation of physically-motivated self-energies typically begin by introducing the injection of states by providing the in-scattering and out-scattering matrices $\Sigma^\text{in}$ and $\Sigma^\text{out}$ respectively. Given these matrices, one can obtain the broadening function $\Gamma$ by directly summing the two: $\Gamma =\Sigma^\text{in} + \Sigma^\text{out}$. To compute the retarded Green's function requires the full self-energy. Fortunately, this can be obtained by noting that the self-energy must be causal. Introducing the Hermitian and anti-Hermitian parts:
\begin{gather}
\left[ \Sigma^r(E) \right]_\text{H} = \frac{1}{2} \left[ \Sigma^r(E) + \Sigma^r(E)^\dagger  \right], \\
 \left[ \Sigma^r(E) \right]_\text{AH} = \frac{1}{2} \left[ \Sigma^r(E) - \Sigma^r(E)^\dagger  \right] = i \Gamma(E)/2
\end{gather}
one can obtain the Hermitian part by the Hilbert transform:
\begin{equation}
\left[ \Sigma^r(E) \right]_\text{H} = \frac{P}{2 \pi i} \int_{-\infty}^{\infty} dE^\prime \frac{ \left[ \Sigma^r(E^\prime) \right]_\text{AH}}{E-E^\prime}. 
\end{equation} 
P denotes the Cauchy principal value:
\begin{equation}
P \int_{-\infty}^{\infty} f(x) dx = \lim_{\epsilon \rightarrow 0^+} \int_{-\infty}^{s - \epsilon} f(x) dx + \int_{s+\epsilon}^{ \infty} f(x) dx
\end{equation}
where $s$ is the location of the singularity. For a constant scalar anti-Hermitian part, as is the case for phenomenological models, the Hilbert transform is identically zero. However more complicated functions tend not to have a fully closed form. For our models we assume two dominant contributions to the scattering, longitudinal-acoustic and longitudinal-optical phonons which we model as B\"uttiker probes and drop the prefix `longitudinal' hereafter. For the following we designate one probe per atomic site, which we enumerate by superscript $p$. For the derivation of the acoustic and optical self-energies, we direct the reader to the work of references \cite{Kubis2011a,Greck2015,Nikonov2011}.

\subsection{Acoustic Phonons}
To arrive at the following self-energy one assumes that the thermal energy is much higher than the phonon energy and that the phonon band being modelled is well-approximated by a linear dispersion $E_\B{q}=v_s |\B{q}|$, where $v_s$ is the speed of sound and $\B{q}$ is the wavevector. Under both these approximations the wavevector dependences cancel and the function is only proportional to the energy value being probed. Additionally, Kubis and Vogl \cite{Kubis2011a} made a final modification to the definition of the self-energy, noting that in the full expression small amounts of energy could be redistributed by small wavevectors, which have now been explicitly disallowed due to the linear dispersion approximation. This deficiency is addressed by inclusion of an energy averaging window from $E-\hbar \omega_D$ to $E+\hbar \omega_D$ where $\omega_D$ is the Debye frequency. We use the same rectangular averaging over energy as Kubis and Vogl, but one could just as easily use any other shaped function, so long as their total integrated area is unity. With these assumptions, the longitudinal-acoustic self-energy is given by
\begin{equation}
\Sigma^{p}_\text{AC} = \frac{1}{\Omega_{p}} \frac{V_D^2 k_B T}{ \rho  v_s^2} \frac{\delta_{\mathbf{r},\mathbf{r}_p}}{2\hbar \omega_D} \int\limits_{E - \hbar \omega_D}^{E +\hbar \omega_D} G^r(E) dE,
\end{equation}
where $\Omega_{p}$ is the elementary volume around point $\B{r}_p$, $\rho$ is the density and $V_D$ is the scalar deformation potential.

\subsection{Optical Phonons} \label{sec:OPphonons}
In this subsection we will detail the model for optical phonons that we use within our B\"uttiker probe method. We begin by discussing the canonical model, the B\"uttiker model as used by Greck \textit{et al.}\cite{Greck2015} and then our approach. In general optical phonons require inter-dependent self-consistent cycles between the retarded Green's function and the electron Green's function. Greck \textit{et al.} \cite{Greck2015} reduced the computational expense of this process by omitting the term related to the electron Green's function in the definition of the retarded self-energy, assuming it to be negligible and using a B\"uttiker probe in place of the in-scattering function. We will now demonstrate that the assumption that the terms related to the electron Green's functions are negligible is contradictory for systems that would be in equilibrium. We address this by amending the approximation to include a new physically motivated non-equilibrium filling function that reduces to the Fermi-Dirac distribution function in equilibrium. 

The canonical dispersion-less optical phonon model is given by the self-energy
\begin{multline}
\Sigma_\text{OP}(E)/D_\text{OP} = \left(n_q + 1\right)  G^r(E-\hbar \omega) + n_q G^r(E+\hbar \omega)  \\ -\frac{i}{2} \left[ G^n(E+\hbar \omega)  - G^n(E-\hbar \omega) \right]  + \\
\frac{P}{2 \pi}\int  \frac{G^n(E^\prime+\hbar \omega)  - G^n(E^\prime-\hbar \omega)}{E-E^\prime} dE^\prime. \label{eq:optphononfull}
\end{multline}
where $D_\text{OP}$ is the scalar electron-phonon coupling and $n_q$ is the Bose-Einstein distribution function for phonon branch $q$,
\begin{equation}
n_q = \left(\exp \left( \frac{\hbar \omega}{k_B T} \right) - 1  \right)^{-1}
\end{equation}
that we have already simplified to use dispersion-less phonons with energy $\hbar \omega$.  Given this expression, both Green's functions are iterated until convergence. A common approximation when modelling optical phonons \cite{Kubis2011a,BMA+2014}, is to remove the Hilbert transform related to the difference between two electron Green's functions from the self-energy in Eq. (\ref{eq:optphononfull}) as is typically assumed to be small. As it is small, the omission of the Hilbert transform in the optical phonon model only slightly affect the observables and current, but by omitting it, the self-energy and hence the retarded Green's function both cease to be causal functions \cite{BMA+2014} and violate the spectral sum rule:
\begin{equation}
\int_{-\infty}^{\infty} A(\B{r},\B{r},E) d\B{r} = 1.
\end{equation}
Historically, for the particular case of optical phonons, the solution is then to do one of two things to restore causality. The first is to actually compute the Hilbert transform, which can be numerically expensive. The second is to omit the difference between the two electron Green's functions entirely while still using the same original in-scattering function. This amounts to making one of two assumptions, either the energy of an optical phonon is assumed to be so low that the Green's function's derivative is practically zero, as done with acoustic phonons, or by implicitly assuming a different out-scattering function that would satisfy these properties. As we typically deal with optical phonon energies many times larger than the thermal energy $k_BT$ it must be the latter. This assumption has an interesting effect. It restores causality and the sum rule but more interestingly decouples the retarded and electron Green's functions. This allows for a considerable reduction in the computation required, as one can converge the retarded Green's function first and then compute the observables at the end. 

Unfortunately, Kubis and Vogl \cite{Kubis2011a} demonstrated that even though such an approximation leads to charge conservation, for certain regimes, \textit{e.g.} in the presence of bound states, this method violates Pauli-blocking, with state occupation numbers exceeding unity. This implies that the assumption about the out-scattering function is unphysical. Only in certain limits where the difference between electron Green's functions in Eq. (\ref{eq:optphononfull}) vanishes identically (such as that of acoustic phonons) can they be decoupled without violating this principle. 

Instead, Greck \textit{et al.} \cite{Greck2015} began from the assumption that the retarded self-energy approximation was physically reasonable but the in/out-scattering functions were now inappropriate. Instead, they treated phonon scattering as energy dependant B\"uttiker probes and calculate fictitious chemical potentials for each lattice site which they called the multiple scattering B\"uttiker probe method \cite{Greck2015}. As the method explicitly uses a Fermi-Dirac distribution function to describe the injection of states, it implicitly obeys conservation rules so long as the probe current is solved to be zero.  In Greck \textit{et al.}'s work, presumably to match the typical form of B\"uttiker probes, only the imaginary component of the retarded Green's function was kept. As explained before, this leads to a violation of causality and we use the full version herein. One could similarly derive models for other dissipation mechanisms such as interfacial roughness or defect scattering but these will not be discussed here.

\subsubsection{Omission of $G^n$ terms}
We begin with the definition of the in-scattering and out-scattering functions for phonons, 
\begin{multline}
\Sigma^\text{in}_\text{OP}(E) = D_\text{OP} n_q G^\text{n}(E-\hbar \omega) \\ +  D_\text{OP} (n_q+1) G^\text{n}(E+\hbar \omega) 
\end{multline}
\begin{multline}
\Sigma^\text{out}_\text{OP}(E) = D_\text{OP} (n_q+1) G^\text{p}(E-\hbar \omega) \\ +  D_\text{OP} n_q G^\text{p}(E+\hbar \omega)
\end{multline}
When the system is in equilibrium, the electron and hole Green's functions $G^\text{n}$ and $G^\text{p}$, can be written as direct products involving the Fermi distribution $f_{eq}$ and the spectral function $A$:
\begin{equation}
G^n(E) = f_\text{eq}(E) A(E), \ G^p(E) =  [1-f_\text{eq}(E)] A(E).
\end{equation}
This allows us to write our in/out-scattering functions as:
\begin{multline}
\Sigma^\text{in}_\text{OP}(E) = D_\text{OP} n_q  f_\text{eq} (E-\hbar \omega) A(E-\hbar \omega) + \\  D_\text{OP} (n_q+1) f_\text{eq}(E+\hbar \omega) A(E+\hbar \omega),
\end{multline}
\begin{multline}
\Sigma^\text{out}_\text{OP}(E) = D_\text{OP} (n_q+1) [1-f_\text{eq}(E-\hbar \omega)]A (E-\hbar \omega) + \\ 
D_\text{OP} n_q [1-f_\text{eq}(E+\hbar \omega)] A(E+\hbar \omega).
\end{multline}
From this, one can then compute the broadening function $\Gamma_\text{ph}(E) = \Sigma^\text{in} + \Sigma^\text{out}$:
\begin{multline}
\Gamma_\text{OP}(E) = D_\text{OP} \left[n_q + 1 - f_\text{eq}(E-\hbar \omega) \right]  A(E-\hbar \omega) \\
+ D_\text{OP} \left[ n_q + f_\text{eq}(E+\hbar \omega) \right] A(E+\hbar \omega). \label{eq:buttikeroptical}
\end{multline}
One can compute the retarded self-energy by using the Hilbert transform. In certain cases, the Hilbert transform of a product of functions can be evaluated as $H[ q(x) r(x)](k)= q(k) H[r(x)](k)$ if the functions satisfy the Bedrosian rules \cite{B1962,HS1998}. We use this as an ansatz to suppose that the Fermi function will not affect the Hilbert transform in an appreciable way, and as the Hilbert transform of the spectral function is analytic, we arrive at the self-energy:
\begin{multline}
\Sigma_\text{OP}(E) = D_\text{OP} \left[n_q + 1 - f_\text{eq}(E-\hbar \omega) \right]  G^r(E-\hbar \omega) + \\ D_\text{OP} \left[ n_q + f_\text{eq}(E+\hbar \omega) \right] G^r(E+\hbar \omega). \label{eq:phSigma} 
\end{multline}
Numerically, for the devices modelled in section \ref{sec:graphenedevices} we found that the approximation of the product deviated less than 5\% at its worst from the full Hilbert transform, and typically did not exceed a fractional error of 1\%. We assume that any other contributions in the retarded self-energy are unimportant. We can now compare this directly to the more commonly cited self-energy in Eq. (\ref{eq:optphononfull}) and by direct comparison to Eqs. (\ref{eq:buttikeroptical}) it is clear that the extra terms proportional to $f_\text{eq}$ correspond to the omitted $G^n$ terms. Now we can make some assessments of this function.

Firstly, let's assume that the system is sufficiently cold that $f_\text{eq}$ can be well approximated by a Heaviside step function, then, for an equilibrium system and all energies $E+\hbar \omega < \mu_\text{eq}$ (where $\mu_\text{eq}$ is the equilibrium chemical potential) then $\Sigma_\text{OP}$ would be exactly given by
\begin{multline}
\Sigma_\text{OP}(E) = D_\text{OP} n_q G(E-\hbar \omega) + \\ D_\text{OP} \left( n_q + 1 \right) G(E+\hbar \omega). \label{eq:eqSigma}
\end{multline}
From this we can see that for this simple equilibrium case, the effect of the correlation terms are not only non-negligible but that they swap the coefficients of the $G(E \pm \hbar \omega)$ terms entirely. When using such a model, absorption processes overly contribute to the total self-energy and emission processes are highly suppressed.

\subsubsection{Compatibility with B\"uttiker probes}
The original goal of B\"uttiker probes for use with optical phonons was to correct for unphysical effects when omitting the $G^n$ terms. One of the worst annoyances for computing the self-energy is the Hilbert transform of the electron Green's function difference. To remedy this, we introduce a new spatially dependent non-equilibrium filling function $f_\text{neq}^p$, which can no longer be described by a simple Fermi function but is still bounded between zero and one:
\begin{equation}
f_\text{neq}^p(\B{r}_p,E) =  G^n(\B{r}_p,\B{r}_p,E)/A(\B{r}_p,\B{r}_p,E).
\end{equation}
We use this function in place of the equilibrium filling function in our earlier derivation, which gives the optical contribution for the $p^{th}$ probe as:
\begin{multline}
\Sigma_\text{OP}^p(E) = \delta_{\B{r},\B{r}_p} D_\text{OP} \left[ n_q + f_\text{neq}^p(E+\hbar \omega) \right] G(E+\hbar \omega) + \\ \delta_{\B{r},\B{r}_p} D_\text{OP} \left[n_q + 1 - f_\text{neq}^p(E-\hbar \omega) \right]  G(E-\hbar \omega).
\end{multline}
The longitudinal-optical phonon self-energy is now a non-equilibrium expression, and requires iteration of $f_\text{neq}$ to converge. Despite no longer being an equilibrium approximation, we still find heuristically that the error between the full Hilbert transform and our approximation did not exceed 3\% and was typically lower than 1\%. 

This self-energy now more accurately takes into account the occupancies of electron states, is compatible with B\"uttiker probes and no longer requires the Hilbert transform. As such it achieves all the positive properties as guaranteed by B\"uttiker probes. Unfortunately minor violation of the spectral sum rule will be observed but we consider our method to be an impressive improvement over the existing approximation as detailed in \cite{Greck2015}. We believe this issue is also further mitigated by $f_\text{neq}$ being a ratio that stays fixed during the self-consistent cycle for $G$, acting as an active update with regards to the change in $G^\text{n}$ unlike the traditional method which remains fixed in magnitude per iteration. This allows us to include optical/acoustic phonons and semi-classical effects all at once. Furthermore, in our experience not only do B\"uttiker probe chemical potentials take less time to compute than that of the older self-consistent method but the probes themselves tend not to change very much per iteration, resulting in them being much closer to convergence and thereby improving the rate of convergence per iteration.

\section{Example: Graphene Nanoribbons}\label{sec:graphenedevices}

\begin{figure}\centering
\includegraphics[width=0.8\columnwidth]{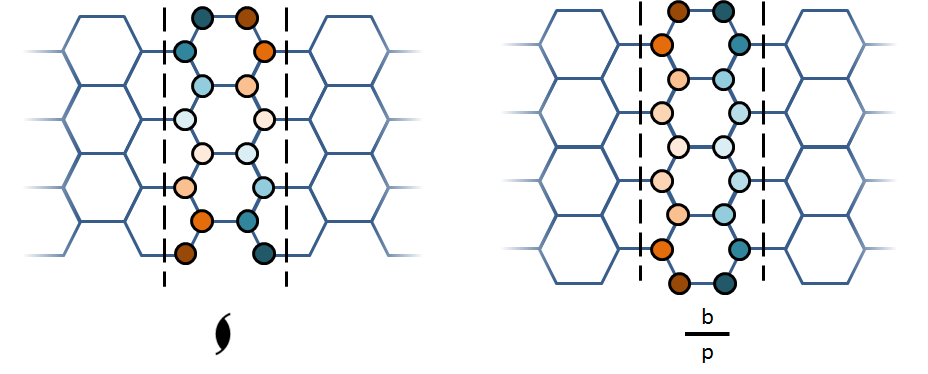}
\caption{Unit cells of armchair nanoribbons, each unit cell consists of $2\times n_y$ carbon atoms. Even $n_y$ configurations have screw symmetry, whereas odd $n_y$ configurations have mirror symmetry. Strength and pigment of colours demonstrate equivalent atoms under (left) screw and (right) mirror operations. Note that while the screw symmetry of even $n_y$ nanoribbons is broken by differing chemical potentials, the mirror symmetry of odd $n_y$ nanoribbons persists. }
\label{fig:armchair}
\end{figure}

A demonstration of the efficacy of our B\"uttiker probe algorithm requires a simple, symmetry free model. An ideal model test case is that of armchair graphene nanoribbons in a fully atomistic representation. For the examples shown here we choose a nanoribbon such that the base nanoribbon itself possesses chiral symmetry (as demonstrated in Fig.~\ref{fig:armchair}), which is then broken by the difference in chemical potentials resulting in a system with no overall symmetry. 

For any system that contains symmetry, one could appropriately set the chemical potentials of symmetric lattice sites to be equal, further minimizing the computational load. For instance, if simulating a device with mirror symmetry, one could set all upper probes to be the same as the lower probes reducing the free parameters by half and reducing the Jacobian size by a factor of 4. Similarly for higher dimensional systems, for example those with cubic symmetry, one could reduce the number of free parameters by 8 and reduce the size of the Jacobian by a factor of 64.

For our calculations we consider only spin-degenerate electrons, though spin-resolved calculations are also completely viable under adequate substitution. The particular simulation parameters and convergence criteria are described in their respective sections. Lead self-energies are solved ahead of time using the Sancho-Rubio/decimation method \cite{SSR1985,ONK+2010} and are reused for subsequent evaluations. Scattering is included in the leads assuming an equilibrium Fermi-Dirac distribution function, and is not updated as this would also contribute to further iterations which are not of interest to our algorithmic discussion. As such, minor effects such as the slight piling of charges appears at the interface but this is of little consequence to the dynamics and dissipation found within the scattering region.

\begin{figure}[t!]\centering
\includegraphics{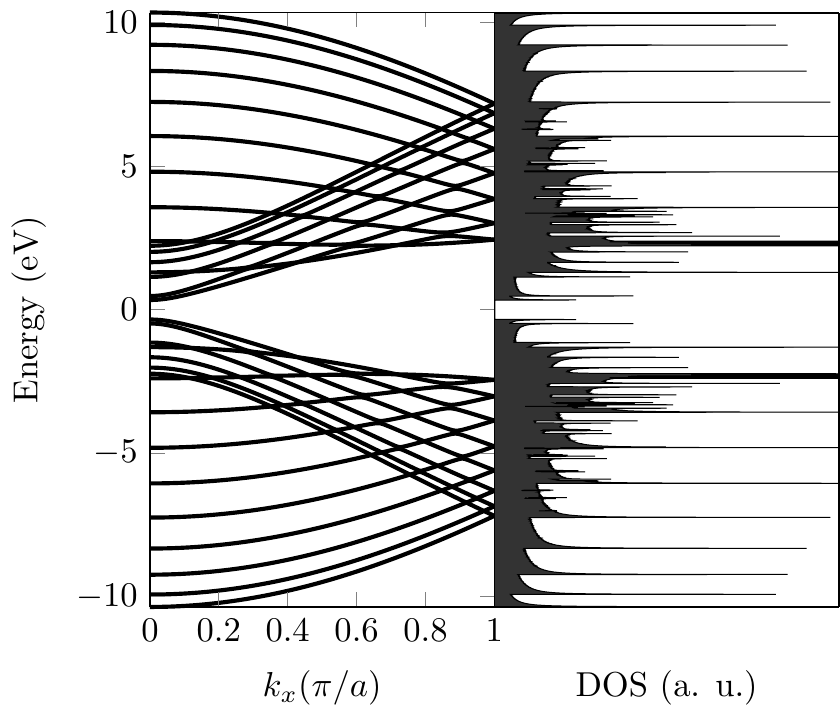}
\caption{(Left) Electronic band structure and (Right) electronic density of states (DOS) for the $n_y=16$ armchair nanoribbon. Only positive $k_x$ are shown as the band structure is symmetric about $k_x=0$. The band gap between the valence and conduction bands is approximately $680$ meV and the separation between the first and second bands at $k_x=0$ is approximately $138$  meV.} 
\label{fig:BANDDOS}
\end{figure}

The Hamiltonian for spin-degenerate graphene can be written in second-quantised form as:
\begin{equation}
H = t_1 \sum\limits_{\langle i,j \rangle} c_i^\dagger c_j +  t_3\! \sum\limits_{\lllangle k,l \rrrangle} c_k^\dagger c_l
\end{equation}
$c \ (c^\dagger)$ is the fermionic annihilation (creation) operator for the $2p_z$ orbitals in graphene, the coefficients are given by $t_1 = -3.2$ eV, and $t_3=-0.3$ eV \cite{White2007}, and the angled bracket subscripts $\langle i,j \rangle, \lllangle k,l \rrrangle$ indicate that the summations are performed over index pairs $i,j$ and $k,l$ that correspond to nearest neighbours and third-nearest neighbours respectively. To provide a simple description, dangling bonds are passivated by hydrogen, whose states are too far away from the energy scales of interest and can be safely ignored \cite{White2007}. No special treatment is given for the edge carbon atoms though it has been suggested that their hopping and on-site terms be modified to adequately represent edge bond distortion \cite{White2007,Grassi2013}. In our model we have included only the first and third hopping terms as both the on-site energy and second-nearest neighbour interactions serve to only rigidly shift the position of the conduction and valence bands and therefore without loss of generality are set to zero. This means that under zero bias, the centre of the band gap is placed exactly at zero \cite{White2007} and we demonstrate this in Fig.~\ref{fig:BANDDOS}. 

As demonstrated in Fig. \ref{fig:armchair}, armchair nanoribbons can be uniquely defined by $2\times n_y$, the number of atoms in their unit cell. Though the typical dynamics of armchair nanoribbons are similar for different $n_y$, their band gaps can drastically differ. For $n_y+1$ divisible by 3, the band gap is incredibly small. For models that do not include third nearest neighbour interactions the band gap is erroneously zero \cite{White2007}.

\subsection{Example devices}
We now provide examples of dissipation in graphene nanoribbon based transistors. Though graphene is well studied, the particular choice of parameters governing the strength of the optical and acoustic contributions is contentious \cite{Yoon2011,Grassi2013,Koswatta2007a,KLN2008}. We choose the numerical coefficients for the optical and acoustic contributions to be $0.07 \text{ eV}^2$ and $0.01 \text{ eV}^2$ respectively, in line with the research of \cite{Yoon2011}. The optical phonon energy $\hbar \omega$ is 180 meV and the Debye energy used is 36 meV. For comparison with previously published results, we assume a temperature of 300 K. As noted in the previous section we choose $n_y=16$ throughout. The central gate regions of our model devices are 30 nm and the total device lengths are 90 nm. 

The convergence criterion for $f_\text{neq}$ is that it changes no more than $10^{-5}$ per iteration, and the B\"uttiker probes are solved such that the current is no greater that $0.1$ pA anywhere. For the following data we use the definition $V_\text{DS}=V_\text{S}-V_\text{D}$ as the voltage difference between the source and drain and $V_\text{GS}=V_\text{G}-V_\text{S}$ as the difference between the gate and source. For convenience and easy comparison to the band structure demonstrated in Fig. \ref{fig:BANDDOS}, the drain voltage is always kept at exactly zero.

\begin{figure}[t!]
\includegraphics[width=\columnwidth]{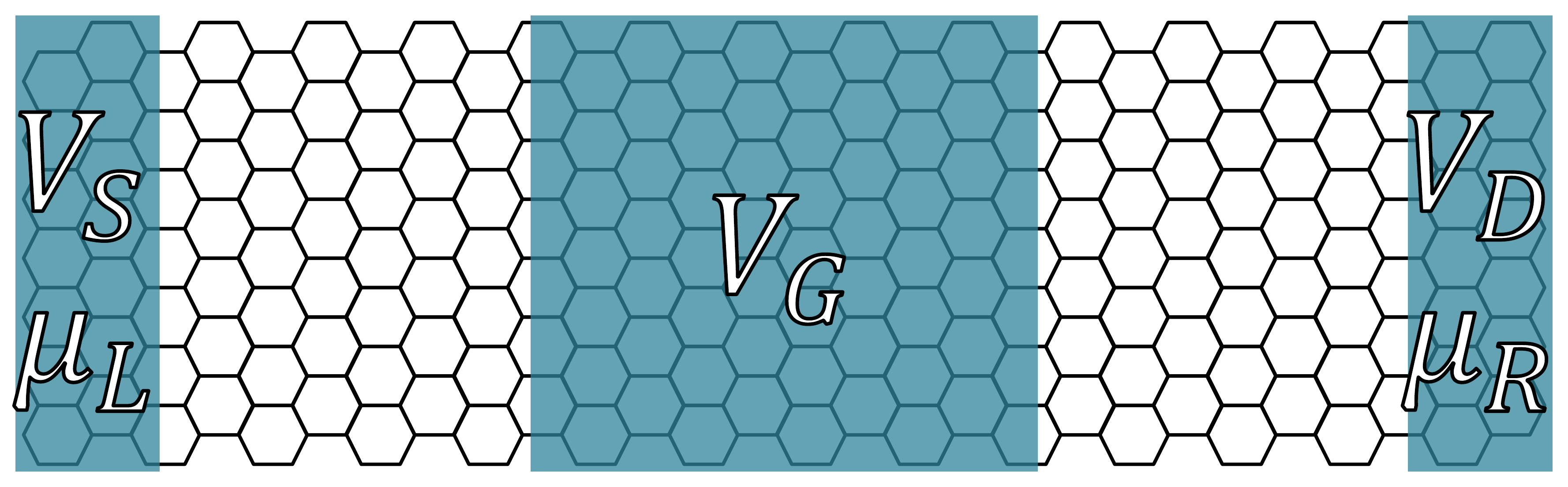}
\caption{A schematic demonstrating the design of the graphene FETS. Three gates, $V_S, V_G, V_D$, are applied to the device which rigidly shift the position of the bands (Fig. \ref{fig:BANDDOS}). $V_S$ and $V_D$ are placed on the semi-infinite contacts with chemical potentials $\mu_\text{S}$, $\mu_\text{D}$ corresponding to the source and drain respectively, and $V_G$ shifts the energy of the central third of the atoms in the device. Intermediate regions then equilibrate by electron-electron interactions\cite{D2005}. The number of atoms in the transverse direction is to scale whereas the longitudinal direction has been reduced to approximately one eighth for ease of viewing. For our simulations we mimic the smooth interpolation between regions by replacing the step functions that describe the gates by error functions.}
\end{figure}

\subsection{Device 1: N-I-N Field effect transistor}
\begin{figure*}[h!]\centering
\includegraphics{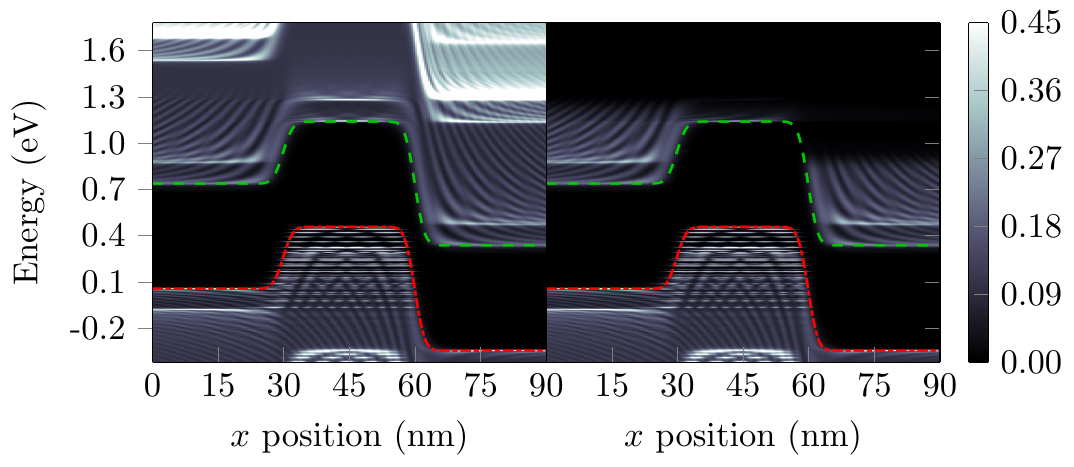}
\includegraphics{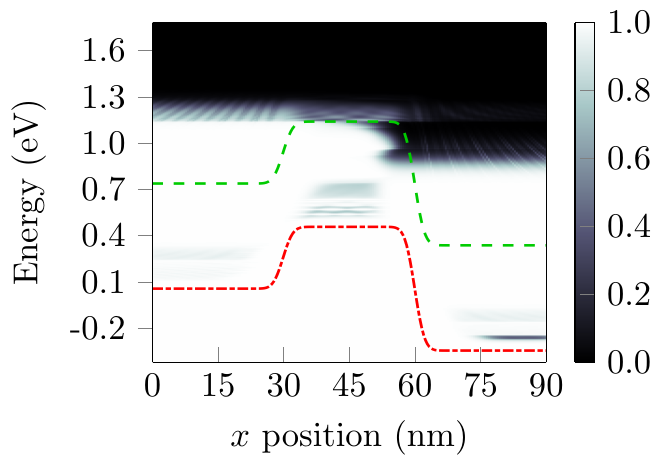}
\caption{Electronic response of a biased NINFET using the RGF method, incorporating both acoustic and optical phonon scattering as a function of position within the device. (Left) Spectral function, $A$. (Middle) Electron Green's function, $G^\text{n}$. (Right) non-equilibrium filling function $f_\text{neq}=G^n/A$. Green dashed line corresponds to the conduction band edge and the red dash-dotted line corresponds to the valence band edge. Both the $A$ and $G^\text{n}$ matrices are in units of eV$^{-1}$, and $f_\text{neq}$ is by definition unitless. Data are averaged over unit cells for presentation.}
\label{fig:converged}
\end{figure*}

\begin{figure*}[h!]\centering
\includegraphics{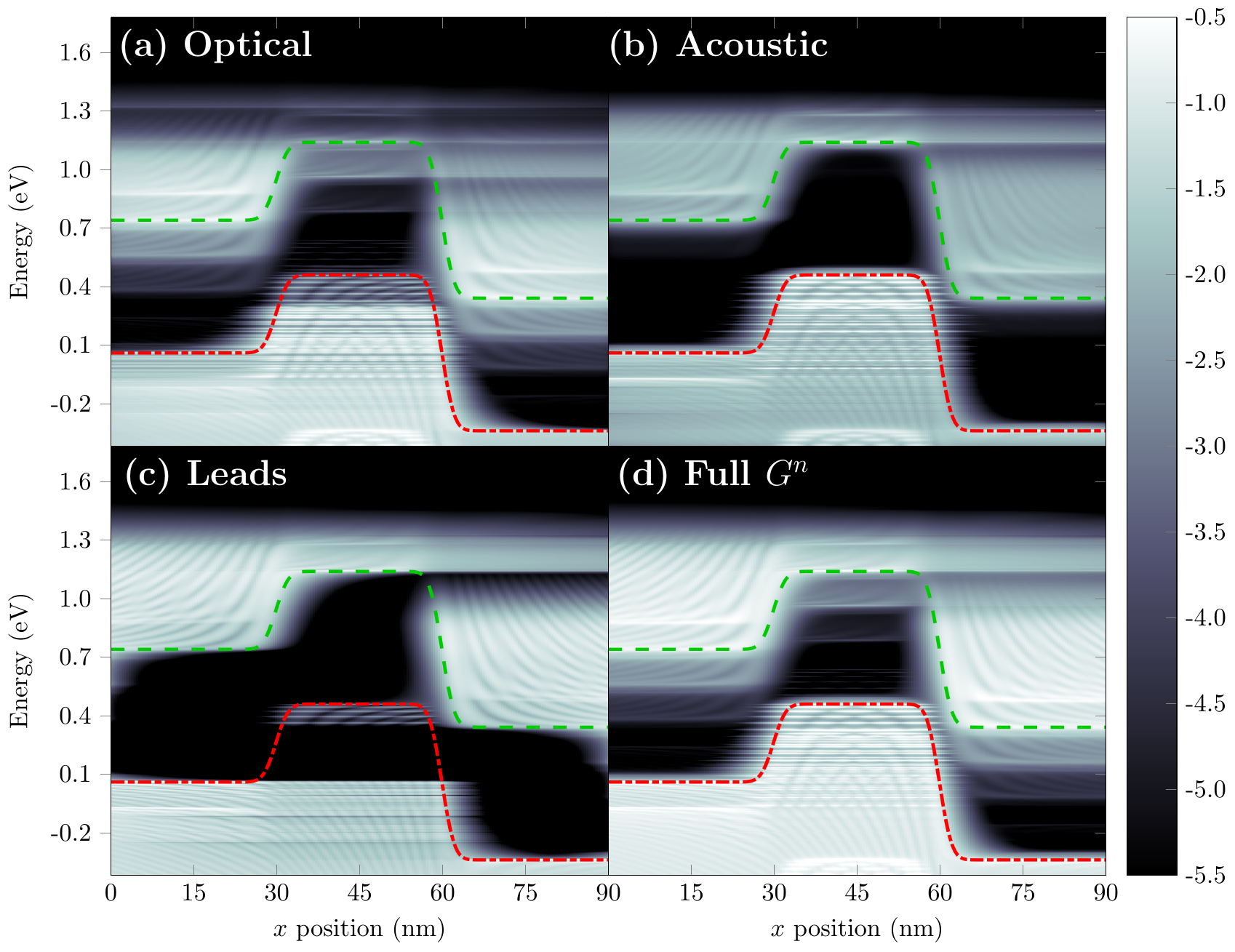}
\caption{NINFET energy resolved partial $G^\text{n}$ matrices in units of eV$^{-1}$. (a) Optical phonon contribution. (b) Acoustic phonon contribution. (c) Lead contributions (including dissipation). (d) All components summed. Green dashed line corresponds to the conduction band edge and the red dash-dotted line corresponds to the valence band edge.  Data are averaged over unit cells and presented using a $\log_{10}$ color axis.}
\label{fig:partialGn}
\end{figure*}

We model a n-i-n field effect transistor (NINFET), the operational parameters are $\mu_\text{S} = 1.240\text{ eV}, \mu_\text{D} = 880\text{ meV}, V_\text{DS}=0.4\text{ V}$, and $V_\text{GS}=0.4\text{ V}$. This model was designed to mimic the devices presented in \cite{Koswatta2007a,Yoon2011}. We present the outputs of these simulations in Figs. \ref{fig:converged},\ref{fig:partialGn}. Convergence of the density of states and electron Green's function is observed in Fig. \ref{fig:converged}. We see that despite our approximation to the self-energy there is no violation of Pauli blocking; no state occupation exceeds unity, as demonstrated in the converged $f_\text{neq}$. To illustrate more clearly the effects of our phonon models, we have plotted the partial charge densities originating from the optical, acoustic and lead self-energies beside the full charge density in Fig. \ref{fig:partialGn}. 

For the optical contribution (Fig.~\ref{fig:partialGn}~a) we see several interesting properties, the most recognisable being the discretely spaced levels easily visible below the conduction band due to phonon emission processes. Due to the splitting between the first and second subbands of graphene being $138$ meV and the optical transition being $180$ meV, one can observe an extremely fine ($42$ meV) level splitting predominantly just below the conduction band and then again one optical transition below it. 

In the barrier region, states are scattered into a level inside the gap, and optical phonon processes are the dominant contribution to states in this region. These optical phonon induced states continue through the right-hand side of the device where few states were filled by the leads. The occupation of these left-moving states is non-zero, as evidenced by the small interference fringes. Not all states are phonon-emission mediated and phonon-absorption states can be observed appearing above the left chemical potential $\mu_\text{S}=1.240$ eV. As these states are isotropically injected, they have little-to-no interference fringes. In general, the proportionality of occupied states by optical transitions appears to be commensurate with the expected behaviour; emission processes are easier, and phonons are emitted or absorbed more strongly when empty or partially empty states are nearby, as evidenced by the non-equilibrium distribution function $f_\text{neq}$ in Fig. \ref{fig:converged}.

The acoustic contribution (Fig.~\ref{fig:partialGn}~b) is a continuous redistribution of states between $E \pm \hbar \omega$ and this is well evidenced by the interference fringes being effectively `washed out'. The most obvious acoustic-based effect is the redistribution of states at $1.15$ eV located between $60$ to $90$ nm; very few of these states were occupied directly by the leads and have been filled mainly by dissipative sources.

As dissipation was included in the leads, the lead self-energy contributions (Fig.~\ref{fig:partialGn}~c) are not ballistic; dissipative effects can be observed. The most noticeable contribution observable is the decaying states directly below the conduction band that are classically forbidden. These states originate from optical/acoustic processes contained in the leads.

\subsection{P-I-N Field effect transistor}

\begin{figure*}[h!]\centering
\includegraphics{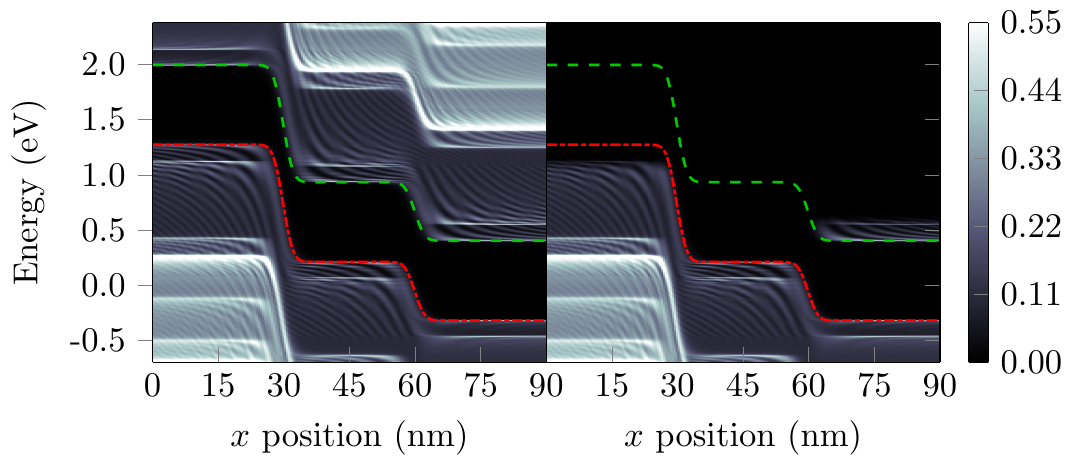}
\includegraphics{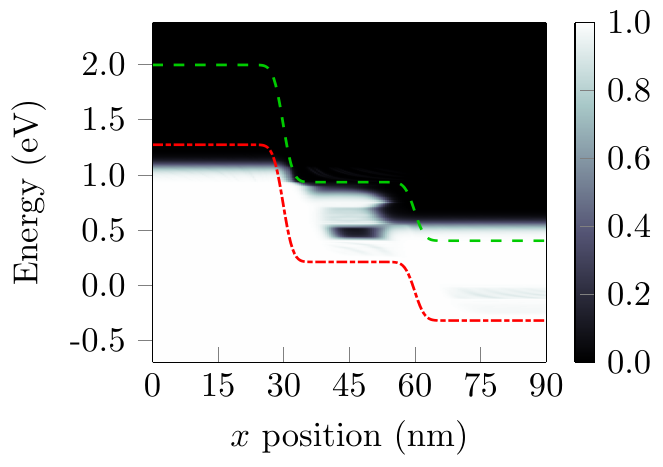}
\caption{Electronic response of a biased PINFET using the RGF method, incorporating both acoustic and optical phonon scattering as a function of position within the device. (Left) Spectral function, $A$. (Middle) Electron Green's function, $G^\text{n}$. (Right) non-equilibrium filling function $f_\text{neq}=G^n/A$. Green dashed line corresponds to the conduction band edge and the red dash-dotted line corresponds to the valence band edge. Both the $A$ and $G^\text{n}$ matrices are in units of eV$^{-1}$, and $f_\text{neq}$ is by definition unitless. Data are averaged over unit cells for presentation.}
\label{fig:PINconverged}
\end{figure*}

\begin{figure*}[h!]\centering
\includegraphics{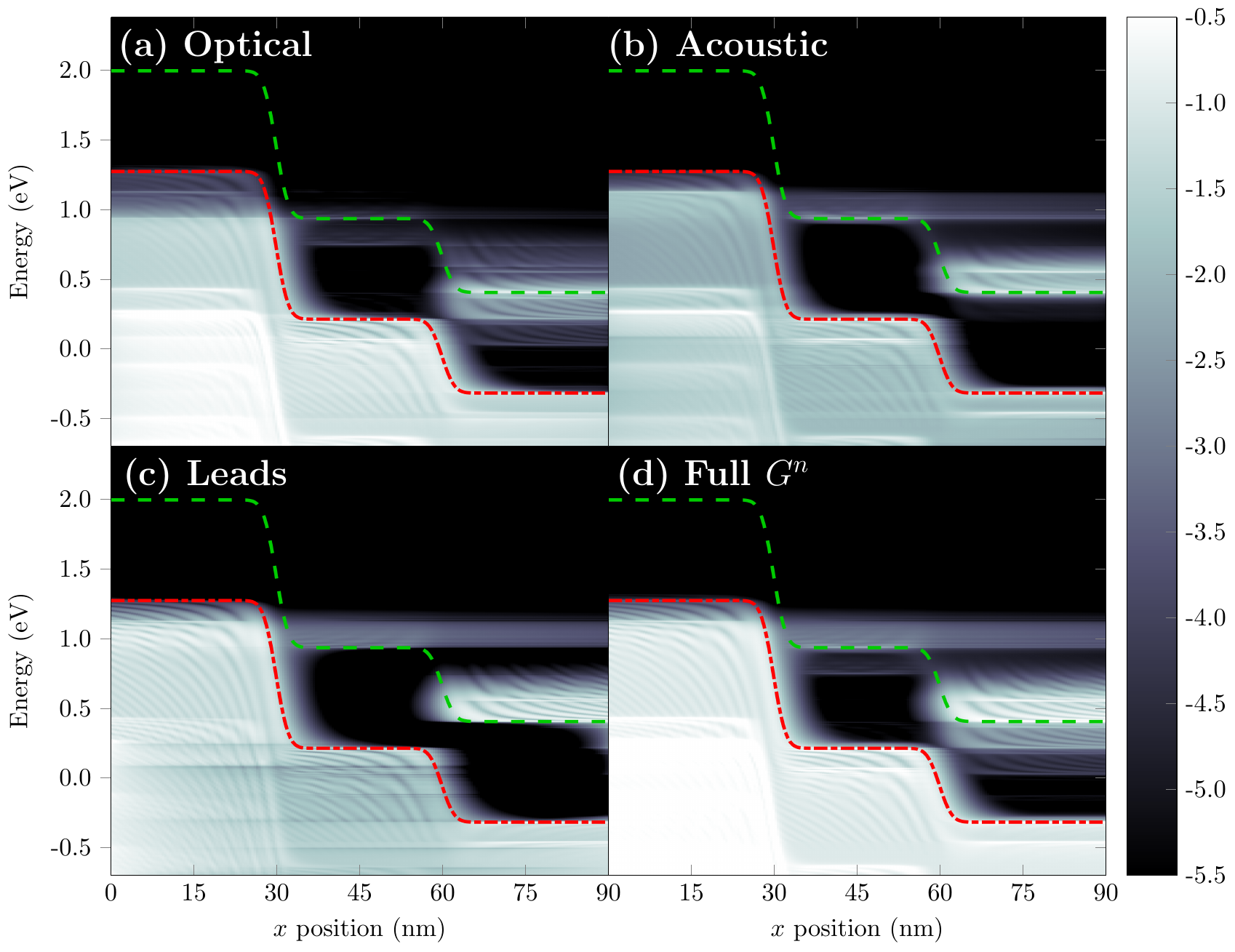}
\caption{PINFET energy resolved partial $G^\text{n}$ matrices in units of eV$^{-1}$. (a) Optical phonon contribution. (b) Acoustic phonon contribution. (c) Lead contributions (including dissipation). (d) All components summed. Green dashed line corresponds to the conduction band edge and the red dash-dotted line corresponds to the valence band edge. Data are averaged over unit cells and presented using a $\log_{10}$ color axis.}
\label{fig:PINpartialGn}
\end{figure*}

We model a p-i-n field effect transistor (PINFET), the operational parameters are: $\mu_\text{S} = 980 \text{meV}, \mu_\text{D} = 480 \text{meV}, V_\text{DS}=1.5\text{V}, V_\text{GS}=-1.0\text{V}$. This device was designed to mimic \cite{KLN2008}.

We present the outputs of these simulations in Figs. \ref{fig:PINconverged},\ref{fig:PINpartialGn} corresponding to the equivalent plots of Figs. \ref{fig:converged},\ref{fig:partialGn} for the NINFET. As before, we see no violation of Pauli blocking; no state occupation exceeds unity, as demonstrated in the converged $f_\text{neq}$.

The optical contribution (Fig.~\ref{fig:PINpartialGn}~a) contains a great deal of fine structure. Beginning from $x=0 \text{ to } 30$nm, an abrupt change in the occupation of states can be observed at approximately $1$eV to the top of the valence band, this width corresponds to exactly two optical phonon emissions $2 \hbar \omega =360$meV. We then see in the region between $30 \text{ to } 60$nm, an appearance of states just below the conduction band emerges. The interference pattern suggest that these states are interfering with left and right-moving states. These optical phonon-based states are the dominant contribution to the occupied states in this region. Unlike the NINFET, these states do not entirely continue onto the right most part of the device, with a clear decrease in state occupation directly adjacent to them. 

In the region of $60 \text{ to } 90$nm, a great deal of fine structure is observed. As in the NINFET, an extremely fine splitting of 42meV results from the interplay of subbands and the optical transition. This splitting is observed at several places, most noticeably at one optical phonon above the conduction band, and then again for one and two optical emissions thereafter. This is not the only overlapping of states and states originating from the left lead can clearly seen to be adjacent to those below the conduction band in the central region.

The acoustic contribution (Fig.~\ref{fig:PINpartialGn}~b) similar to that in the NINFET again provides a `washing out' of the states. The most obvious acoustic-based effect is the redistribution of states at $750$ meV located between $60$ to $90$ nm; practically none of these states were occupied directly by the leads and have been filled mainly by dissipative sources.

The lead self-energy contributions (Fig.~\ref{fig:PINpartialGn}~c) again similar to the NINFET are not ballistic and classically forbidden decaying states directly below the conduction band arise from the optical/acoustic processes contained in the leads. Minor `streaks' can be observed where there are no states and these are regions that are particularly sensitive to the dissipation mechanisms. This is most clearly evidenced in the region of $60 \text{ to } 90$nm where these states are being being removed for re-injection by optical phonon-based processes.

All of the effects observed using our new approach are comparable to those seen in the literature using more computationally expensive approaches \cite{Koswatta2007a,Yoon2011,KLN2008}.

\section{Conclusion}
An efficient algorithm for the inclusion of dissipation in electronic systems has been presented. The algorithm utilised two well-known algorithms that exist in the literature, namely RGF and B\"uttiker probes. Sparsity properties and various limits were used to greatly improve the runtime of these algorithms with a reduction in operations from $\mathcal{O}(N^3)$ to $\mathcal{O}(N^2)$ for the evaluation of the trace and a reduction of $N_E-1$ evaluations of RGF in the calculation of the Jacobian. These methods were validated by simulation of fully atomistic graphene nanotransistors in n-i-n and p-i-n configurations. The results demonstrated a clear resilience to the approximations made to the Jacobian and the B\"uttiker phonon model very clearly gave effects directly comparable with that of a full first principles approach.


\section{Appendix}

\subsection{$G^n$ with Block-Tridiagonal self-energies}\label{sec:Gn}
It is typically assumed that the self-energies of interest are block-diagonal. In this appendix we provide the algorithm to calculate the recursive Green's function with block-tridiagonal matrices. The algorithm for the retarded Green's function is unchanged from Anantram \cite{ALN2008} as it already assumes block-tridiagonal structure, for the electron Green's function, we introduce the left-connected electron green's function
\begin{multline}
g^{nLi+1}_{i+1,i+1} = g^{rLi}_{i+1,i+1} \left( \Sigma^\text{in}_{i+1,i+1} +  \sigma^\text{in}_{i+1,i+1}  \right. \\
 \left. - \Sigma^\text{in}_{i+1,i} g^{aLi}_{i,i} K^\dagger_{i,i+1} - K_{i+1,i} g^{rLi}_{i,i} \Sigma^\text{in}_{i,i+1} \right) g^{aLi}_{i+1,i+1} \label{eq:fullleft}
\end{multline}
$\sigma^\text{in}_{i+1,i+1} = K_{i+1,i}g^{nLi}_{i,i}K_{i,i+1}^\dagger$, where the $g^{n0}$ matrices are given by:
\begin{align}
g^{n0}_{i,i+1} =& g^{r0}_{i,i} \Sigma^\text{in}_{i,i+1} g^{a0}_{i+1,i+1} \\
g^{n0}_{i+1,i} =& g^{r0}_{i+1,i+1} \Sigma^\text{in}_{i+1,i} g^{a0}_{i,i}
\end{align}
seeding $g^{nL1}_{1,1}= g^{rL1}_{1,1} \Sigma^\text{in}_{1,1} g^{aL1}_{1,1}$ and iterating (\ref{eq:fullleft}) from $i=1$ to $N-1$. Once all left-connected Green's functions are calculated, seeding $G^n_{N,N} = g^{nLN}_{N,N}$ one then finds by backwards recursion:
\begin{multline}
G^{n}_{i+1,i} = g^{n0}_{i+1,i} - G_{i+1,i}K_{i,i+1}g^{n0}_{i+1,i} \\ -G_{i+1,i+1}K_{i+1,i} g^{nLi}_{i,i} - G^{n}_{i+1,i+1} K^\dagger_{i,i+1} g^{aLi}_{i,i},
\end{multline}
$G^n_{i,i+1} = (G^{n}_{i+1,i})^\dagger$, and
\begin{multline}
G^{n}_{i,i} = g^{nLi}_{i,i} - g^{nLi}_{i,i}K^\dagger_{i,i+1} G^\dagger_{i+1,i} \\ - g^{n0}_{i,i+1}K^\dagger_{i+1,i} G^\dagger_{i,i} - g^{rLi}_{i,i} K_{q,q+1} G^{n}_{i+1,i}.
\end{multline}
\ 
\section{Acknowledgements}
The authors would like to thank Jackson S. Smith for fruitful conversations. This work was supported in part by the Australian Research Council under the Discovery and Centre of Excellence funding schemes (project numbers: DP140100375 and CE170100039). Computational resources were provided by the NCI National Facility systems at the Australian National University through the National Computational Merit Allocation Scheme supported by the Australian Government.

%



\begin{thebibliography}{29}%
\makeatletter
\providecommand \@ifxundefined [1]{%
 \@ifx{#1\undefined}
}%
\providecommand \@ifnum [1]{%
 \ifnum #1\expandafter \@firstoftwo
 \else \expandafter \@secondoftwo
 \fi
}%
\providecommand \@ifx [1]{%
 \ifx #1\expandafter \@firstoftwo
 \else \expandafter \@secondoftwo
 \fi
}%
\providecommand \natexlab [1]{#1}%
\providecommand \enquote  [1]{``#1''}%
\providecommand \bibnamefont  [1]{#1}%
\providecommand \bibfnamefont [1]{#1}%
\providecommand \citenamefont [1]{#1}%
\providecommand \href@noop [0]{\@secondoftwo}%
\providecommand \href [0]{\begingroup \@sanitize@url \@href}%
\providecommand \@href[1]{\@@startlink{#1}\@@href}%
\providecommand \@@href[1]{\endgroup#1\@@endlink}%
\providecommand \@sanitize@url [0]{\catcode `\\12\catcode `\$12\catcode
  `\&12\catcode `\#12\catcode `\^12\catcode `\_12\catcode `\%12\relax}%
\providecommand \@@startlink[1]{}%
\providecommand \@@endlink[0]{}%
\providecommand \url  [0]{\begingroup\@sanitize@url \@url }%
\providecommand \@url [1]{\endgroup\@href {#1}{\urlprefix }}%
\providecommand \urlprefix  [0]{URL }%
\providecommand \Eprint [0]{\href }%
\providecommand \doibase [0]{http://dx.doi.org/}%
\providecommand \selectlanguage [0]{\@gobble}%
\providecommand \bibinfo  [0]{\@secondoftwo}%
\providecommand \bibfield  [0]{\@secondoftwo}%
\providecommand \translation [1]{[#1]}%
\providecommand \BibitemOpen [0]{}%
\providecommand \bibitemStop [0]{}%
\providecommand \bibitemNoStop [0]{.\EOS\space}%
\providecommand \EOS [0]{\spacefactor3000\relax}%
\providecommand \BibitemShut  [1]{\csname bibitem#1\endcsname}%
\let\auto@bib@innerbib\@empty
\bibitem [{\citenamefont {Anantram}\ \emph {et~al.}(2008)\citenamefont
  {Anantram}, \citenamefont {Lundstrom},\ and\ \citenamefont
  {Nikonov}}]{ALN2008}%
  \BibitemOpen
  \bibfield  {author} {\bibinfo {author} {\bibfnamefont {M.~P.}\ \bibnamefont
  {Anantram}}, \bibinfo {author} {\bibfnamefont {M.~S.}\ \bibnamefont
  {Lundstrom}}, \ and\ \bibinfo {author} {\bibfnamefont {D.~E.}\ \bibnamefont
  {Nikonov}},\ }\href {\doibase 10.1109/JPROC.2008.927355} {\bibfield
  {journal} {\bibinfo  {journal} {Proc. IEEE}\ }\textbf {\bibinfo {volume}
  {96}},\ \bibinfo {pages} {1511} (\bibinfo {year} {2008})}\BibitemShut
  {NoStop}%
\bibitem [{\citenamefont {Datta}(2005)}]{D2005}%
  \BibitemOpen
  \bibfield  {author} {\bibinfo {author} {\bibfnamefont {S.}~\bibnamefont
  {Datta}},\ }\href {\doibase 10.1017/CBO9781139164313} {\emph {\bibinfo
  {title} {{Quantum Transport: Atom to Transistor}}}}\ (\bibinfo  {publisher}
  {Cambridge University Press},\ \bibinfo {year} {2005})\BibitemShut {NoStop}%
\bibitem [{\citenamefont {Li}\ and\ \citenamefont {Darve}(2012)}]{LD2012}%
  \BibitemOpen
  \bibfield  {author} {\bibinfo {author} {\bibfnamefont {S.}~\bibnamefont
  {Li}}\ and\ \bibinfo {author} {\bibfnamefont {E.}~\bibnamefont {Darve}},\
  }\href {\doibase 10.1016/j.jcp.2011.05.027} {\bibfield  {journal} {\bibinfo
  {journal} {J. Comput. Phys.}\ }\textbf {\bibinfo {volume} {231}},\ \bibinfo
  {pages} {1121} (\bibinfo {year} {2012})}\BibitemShut {NoStop}%
\bibitem [{\citenamefont {Li}\ \emph {et~al.}(2013)\citenamefont {Li},
  \citenamefont {Wu},\ and\ \citenamefont {Darve}}]{LWD2013}%
  \BibitemOpen
  \bibfield  {author} {\bibinfo {author} {\bibfnamefont {S.}~\bibnamefont
  {Li}}, \bibinfo {author} {\bibfnamefont {W.}~\bibnamefont {Wu}}, \ and\
  \bibinfo {author} {\bibfnamefont {E.}~\bibnamefont {Darve}},\ }\href
  {\doibase 10.1016/j.jcp.2013.01.036} {\bibfield  {journal} {\bibinfo
  {journal} {J. Comput. Phys.}\ }\textbf {\bibinfo {volume} {242}},\ \bibinfo
  {pages} {915} (\bibinfo {year} {2013})}\BibitemShut {NoStop}%
\bibitem [{\citenamefont {Polizzi}\ and\ \citenamefont {Sameh}(2006)}]{PS2006}%
  \BibitemOpen
  \bibfield  {author} {\bibinfo {author} {\bibfnamefont {E.}~\bibnamefont
  {Polizzi}}\ and\ \bibinfo {author} {\bibfnamefont {A.~H.}\ \bibnamefont
  {Sameh}},\ }\href {\doibase 10.1016/j.parco.2005.07.005} {\bibfield
  {journal} {\bibinfo  {journal} {Parallel Comput.}\ }\textbf {\bibinfo
  {volume} {32}},\ \bibinfo {pages} {177} (\bibinfo {year} {2006})}\BibitemShut
  {NoStop}%
\bibitem [{\citenamefont {Mamaluy}\ \emph {et~al.}(2005)\citenamefont
  {Mamaluy}, \citenamefont {Vasileska}, \citenamefont {Sabathil}, \citenamefont
  {Zibold},\ and\ \citenamefont {Vogl}}]{MVM+2005}%
  \BibitemOpen
  \bibfield  {author} {\bibinfo {author} {\bibfnamefont {D.}~\bibnamefont
  {Mamaluy}}, \bibinfo {author} {\bibfnamefont {D.}~\bibnamefont {Vasileska}},
  \bibinfo {author} {\bibfnamefont {M.}~\bibnamefont {Sabathil}}, \bibinfo
  {author} {\bibfnamefont {T.}~\bibnamefont {Zibold}}, \ and\ \bibinfo {author}
  {\bibfnamefont {P.}~\bibnamefont {Vogl}},\ }\href {\doibase
  10.1103/PhysRevB.71.245321} {\bibfield  {journal} {\bibinfo  {journal} {Phys.
  Rev. B}\ }\textbf {\bibinfo {volume} {71}},\ \bibinfo {pages} {1} (\bibinfo
  {year} {2005})}\BibitemShut {NoStop}%
\bibitem [{\citenamefont {Chen}\ \emph {et~al.}(2015)\citenamefont {Chen},
  \citenamefont {Li}, \citenamefont {Yam}, \citenamefont {Zhang}, \citenamefont
  {Wong},\ and\ \citenamefont {Chen}}]{CLY+2015}%
  \BibitemOpen
  \bibfield  {author} {\bibinfo {author} {\bibfnamefont {Q.}~\bibnamefont
  {Chen}}, \bibinfo {author} {\bibfnamefont {J.}~\bibnamefont {Li}}, \bibinfo
  {author} {\bibfnamefont {C.}~\bibnamefont {Yam}}, \bibinfo {author}
  {\bibfnamefont {Y.}~\bibnamefont {Zhang}}, \bibinfo {author} {\bibfnamefont
  {N.}~\bibnamefont {Wong}}, \ and\ \bibinfo {author} {\bibfnamefont
  {G.}~\bibnamefont {Chen}},\ }\href {\doibase 10.1016/j.jcp.2015.01.032}
  {\bibfield  {journal} {\bibinfo  {journal} {J. Comput. Phys.}\ }\textbf
  {\bibinfo {volume} {286}},\ \bibinfo {pages} {49} (\bibinfo {year}
  {2015})}\BibitemShut {NoStop}%
\bibitem [{\citenamefont {Venugopal}\ \emph {et~al.}(2003)\citenamefont
  {Venugopal}, \citenamefont {Paulsson}, \citenamefont {Goasguen},
  \citenamefont {Datta},\ and\ \citenamefont {Lundstrom}}]{VPG+2003}%
  \BibitemOpen
  \bibfield  {author} {\bibinfo {author} {\bibfnamefont {R.}~\bibnamefont
  {Venugopal}}, \bibinfo {author} {\bibfnamefont {M.}~\bibnamefont {Paulsson}},
  \bibinfo {author} {\bibfnamefont {S.}~\bibnamefont {Goasguen}}, \bibinfo
  {author} {\bibfnamefont {S.}~\bibnamefont {Datta}}, \ and\ \bibinfo {author}
  {\bibfnamefont {M.~S.}\ \bibnamefont {Lundstrom}},\ }\href {\doibase
  10.1063/1.1563298} {\bibfield  {journal} {\bibinfo  {journal} {J. Appl.
  Phys.}\ }\textbf {\bibinfo {volume} {93}},\ \bibinfo {pages} {5613} (\bibinfo
  {year} {2003})}\BibitemShut {NoStop}%
\bibitem [{\citenamefont {Lake}\ \emph {et~al.}(1997)\citenamefont {Lake},
  \citenamefont {Klimeck}, \citenamefont {Bowen},\ and\ \citenamefont
  {Jovanovic}}]{Lake1997}%
  \BibitemOpen
  \bibfield  {author} {\bibinfo {author} {\bibfnamefont {R.}~\bibnamefont
  {Lake}}, \bibinfo {author} {\bibfnamefont {G.}~\bibnamefont {Klimeck}},
  \bibinfo {author} {\bibfnamefont {R.~C.}\ \bibnamefont {Bowen}}, \ and\
  \bibinfo {author} {\bibfnamefont {D.}~\bibnamefont {Jovanovic}},\ }\href
  {\doibase 10.1063/1.365394} {\bibfield  {journal} {\bibinfo  {journal} {J.
  Appl. Phys.}\ }\textbf {\bibinfo {volume} {81}},\ \bibinfo {pages} {7845}
  (\bibinfo {year} {1997})}\BibitemShut {NoStop}%
\bibitem [{\citenamefont {Nikonov}\ \emph {et~al.}(2011)\citenamefont
  {Nikonov}, \citenamefont {Bourianoff}, \citenamefont {Gargini},\ and\
  \citenamefont {Pal}}]{Nikonov2011}%
  \BibitemOpen
  \bibfield  {author} {\bibinfo {author} {\bibfnamefont {D.}~\bibnamefont
  {Nikonov}}, \bibinfo {author} {\bibfnamefont {G.}~\bibnamefont {Bourianoff}},
  \bibinfo {author} {\bibfnamefont {P.}~\bibnamefont {Gargini}}, \ and\
  \bibinfo {author} {\bibfnamefont {H.}~\bibnamefont {Pal}},\ }\href
  {https://nanohub.org/resources/7772} {\enquote {\bibinfo {title} {{Scattering
  in NEGF : Made simple}},}\ } (\bibinfo {year} {2011})\BibitemShut {NoStop}%
\bibitem [{\citenamefont {Greck}\ \emph {et~al.}(2015)\citenamefont {Greck},
  \citenamefont {Schindler},\ and\ \citenamefont {Vogl}}]{Greck2015}%
  \BibitemOpen
  \bibfield  {author} {\bibinfo {author} {\bibfnamefont {P.}~\bibnamefont
  {Greck}}, \bibinfo {author} {\bibfnamefont {C.}~\bibnamefont {Schindler}}, \
  and\ \bibinfo {author} {\bibfnamefont {P.}~\bibnamefont {Vogl}},\ }\href
  {\doibase 10.1364/OE.23.006587} {\bibfield  {journal} {\bibinfo  {journal}
  {Opt. Express}\ }\textbf {\bibinfo {volume} {23}},\ \bibinfo {pages} {6587}
  (\bibinfo {year} {2015})}\BibitemShut {NoStop}%
\bibitem [{\citenamefont {Sadasivam}\ \emph {et~al.}(2017)\citenamefont
  {Sadasivam}, \citenamefont {Ye}, \citenamefont {Feser}, \citenamefont
  {Charles}, \citenamefont {Miao}, \citenamefont {Kubis},\ and\ \citenamefont
  {Fisher}}]{Sadasivam2017}%
  \BibitemOpen
  \bibfield  {author} {\bibinfo {author} {\bibfnamefont {S.}~\bibnamefont
  {Sadasivam}}, \bibinfo {author} {\bibfnamefont {N.}~\bibnamefont {Ye}},
  \bibinfo {author} {\bibfnamefont {J.~P.}\ \bibnamefont {Feser}}, \bibinfo
  {author} {\bibfnamefont {J.}~\bibnamefont {Charles}}, \bibinfo {author}
  {\bibfnamefont {K.}~\bibnamefont {Miao}}, \bibinfo {author} {\bibfnamefont
  {T.}~\bibnamefont {Kubis}}, \ and\ \bibinfo {author} {\bibfnamefont {T.~S.}\
  \bibnamefont {Fisher}},\ }\href {\doibase 10.1103/PhysRevB.95.085310}
  {\bibfield  {journal} {\bibinfo  {journal} {Phys. Rev. B}\ }\textbf {\bibinfo
  {volume} {95}},\ \bibinfo {pages} {085310} (\bibinfo {year}
  {2017})}\BibitemShut {NoStop}%
\bibitem [{\citenamefont {Wimmer}\ and\ \citenamefont
  {Richter}(2009)}]{Wimmer2009}%
  \BibitemOpen
  \bibfield  {author} {\bibinfo {author} {\bibfnamefont {M.}~\bibnamefont
  {Wimmer}}\ and\ \bibinfo {author} {\bibfnamefont {K.}~\bibnamefont
  {Richter}},\ }\href {\doibase 10.1016/j.jcp.2009.08.001} {\bibfield
  {journal} {\bibinfo  {journal} {J. Comput. Phys.}\ }\textbf {\bibinfo
  {volume} {228}},\ \bibinfo {pages} {8548} (\bibinfo {year}
  {2009})}\BibitemShut {NoStop}%
\bibitem [{\citenamefont {Eyert}(1996)}]{Eyert1996}%
  \BibitemOpen
  \bibfield  {author} {\bibinfo {author} {\bibfnamefont {V.}~\bibnamefont
  {Eyert}},\ }\href {\doibase 10.1006/jcph.1996.0059} {\bibfield  {journal}
  {\bibinfo  {journal} {J. Comput. Phys.}\ }\textbf {\bibinfo {volume} {124}},\
  \bibinfo {pages} {271} (\bibinfo {year} {1996})}\BibitemShut {NoStop}%
\bibitem [{\citenamefont {Harris}\ and\ \citenamefont
  {Stöcker}(1998)}]{HS1998}%
  \BibitemOpen
  \bibfield  {author} {\bibinfo {author} {\bibfnamefont {J.~W.}\ \bibnamefont
  {Harris}}\ and\ \bibinfo {author} {\bibfnamefont {H.}~\bibnamefont
  {Stöcker}},\ }\href
  {https://books.google.com.au/books?hl=en{\&}lr={\&}id=DnKLkOb{\_}YfIC{\&}oi=fnd{\&}pg=PR5{\&}dq=harris+mathematics{\&}ots=vub7EmOn{\_}6{\&}sig=C-i9O5DJOeFOWCXqJ6U7xzinIBo{\#}v=onepage{\&}q=harris
  mathematics{\&}f=false} {\emph {\bibinfo {title} {{Handbook of mathematics
  and computational science}}}}\ (\bibinfo  {publisher} {Springer},\ \bibinfo
  {year} {1998})\BibitemShut {NoStop}%
\bibitem [{\citenamefont {Levenberg}(1944)}]{L1944}%
  \BibitemOpen
  \bibfield  {author} {\bibinfo {author} {\bibfnamefont {K.}~\bibnamefont
  {Levenberg}},\ }\href {\doibase 10.1090/qam/10666} {\bibfield  {journal}
  {\bibinfo  {journal} {Q. Appl. Math.}\ }\textbf {\bibinfo {volume} {2}},\
  \bibinfo {pages} {164} (\bibinfo {year} {1944})}\BibitemShut {NoStop}%
\bibitem [{\citenamefont {Marquardt}(1963)}]{M1963}%
  \BibitemOpen
  \bibfield  {author} {\bibinfo {author} {\bibfnamefont {D.~W.}\ \bibnamefont
  {Marquardt}},\ }\href {\doibase 10.1137/0111030} {\bibfield  {journal}
  {\bibinfo  {journal} {J. Soc. Ind. Appl. Math.}\ }\textbf {\bibinfo {volume}
  {11}},\ \bibinfo {pages} {431} (\bibinfo {year} {1963})}\BibitemShut
  {NoStop}%
\bibitem [{\citenamefont {Steihaug}(1983)}]{S1983}%
  \BibitemOpen
  \bibfield  {author} {\bibinfo {author} {\bibfnamefont {T.}~\bibnamefont
  {Steihaug}},\ }\href {\doibase 10.1137/0720042} {\bibfield  {journal}
  {\bibinfo  {journal} {SIAM J. Numer. Anal.}\ }\textbf {\bibinfo {volume}
  {20}},\ \bibinfo {pages} {626} (\bibinfo {year} {1983})}\BibitemShut
  {NoStop}%
\bibitem [{\citenamefont {Fletcher}\ and\ \citenamefont
  {Powell}(1963)}]{FP1963}%
  \BibitemOpen
  \bibfield  {author} {\bibinfo {author} {\bibfnamefont {R.}~\bibnamefont
  {Fletcher}}\ and\ \bibinfo {author} {\bibfnamefont {M.~J.~D.}\ \bibnamefont
  {Powell}},\ }\href {\doibase 10.1093/comjnl/6.2.163} {\bibfield  {journal}
  {\bibinfo  {journal} {Comput. J.}\ }\textbf {\bibinfo {volume} {6}},\
  \bibinfo {pages} {163} (\bibinfo {year} {1963})}\BibitemShut {NoStop}%
\bibitem [{\citenamefont {Kubis}\ and\ \citenamefont
  {Vogl}(2011)}]{Kubis2011a}%
  \BibitemOpen
  \bibfield  {author} {\bibinfo {author} {\bibfnamefont {T.}~\bibnamefont
  {Kubis}}\ and\ \bibinfo {author} {\bibfnamefont {P.}~\bibnamefont {Vogl}},\
  }\href {\doibase 10.1103/PhysRevB.83.195304} {\bibfield  {journal} {\bibinfo
  {journal} {Phys. Rev. B}\ }\textbf {\bibinfo {volume} {83}},\ \bibinfo
  {pages} {1} (\bibinfo {year} {2011})}\BibitemShut {NoStop}%
\bibitem [{\citenamefont {Barker}\ \emph {et~al.}(2014)\citenamefont {Barker},
  \citenamefont {Martinez}, \citenamefont {Aldegunde},\ and\ \citenamefont
  {Valin}}]{BMA+2014}%
  \BibitemOpen
  \bibfield  {author} {\bibinfo {author} {\bibfnamefont {J.~R.}\ \bibnamefont
  {Barker}}, \bibinfo {author} {\bibfnamefont {A.}~\bibnamefont {Martinez}},
  \bibinfo {author} {\bibfnamefont {M.}~\bibnamefont {Aldegunde}}, \ and\
  \bibinfo {author} {\bibfnamefont {R.}~\bibnamefont {Valin}},\ }\href
  {\doibase 10.1088/1742-6596/526/1/012001} {\bibfield  {journal} {\bibinfo
  {journal} {J. Phys. Conf. Ser.}\ }\textbf {\bibinfo {volume} {526}},\
  \bibinfo {pages} {012001} (\bibinfo {year} {2014})}\BibitemShut {NoStop}%
\bibitem [{\citenamefont {Bedrosian}(1962)}]{B1962}%
  \BibitemOpen
  \bibfield  {author} {\bibinfo {author} {\bibfnamefont {E.}~\bibnamefont
  {Bedrosian}},\ }\href
  {http://www.rand.org/pubs/research{\_}memoranda/RM3439.html} {\emph {\bibinfo
  {title} {{A Product Theorem for Hilbert Transforms}}}},\ \bibinfo {type}
  {Tech. Rep.}\ (\bibinfo {year} {1962})\BibitemShut {NoStop}%
\bibitem [{\citenamefont {Sancho}\ \emph {et~al.}(1985)\citenamefont {Sancho},
  \citenamefont {Sancho},\ and\ \citenamefont {Rubio}}]{SSR1985}%
  \BibitemOpen
  \bibfield  {author} {\bibinfo {author} {\bibfnamefont {M.~P.~L.}\
  \bibnamefont {Sancho}}, \bibinfo {author} {\bibfnamefont {J.~M.~L.}\
  \bibnamefont {Sancho}}, \ and\ \bibinfo {author} {\bibfnamefont
  {J.}~\bibnamefont {Rubio}},\ }\href {\doibase 10.1088/0305-4608/15/4/009}
  {\bibfield  {journal} {\bibinfo  {journal} {J. Phys. F Met. Phys.}\ }\textbf
  {\bibinfo {volume} {15}},\ \bibinfo {pages} {851} (\bibinfo {year}
  {1985})}\BibitemShut {NoStop}%
\bibitem [{\citenamefont {Ozaki}\ \emph {et~al.}(2010)\citenamefont {Ozaki},
  \citenamefont {Nishio},\ and\ \citenamefont {Kino}}]{ONK+2010}%
  \BibitemOpen
  \bibfield  {author} {\bibinfo {author} {\bibfnamefont {T.}~\bibnamefont
  {Ozaki}}, \bibinfo {author} {\bibfnamefont {K.}~\bibnamefont {Nishio}}, \
  and\ \bibinfo {author} {\bibfnamefont {H.}~\bibnamefont {Kino}},\ }\href
  {\doibase 10.1103/PhysRevB.81.035116} {\bibfield  {journal} {\bibinfo
  {journal} {Phys. Rev. B}\ }\textbf {\bibinfo {volume} {81}},\ \bibinfo
  {pages} {035116} (\bibinfo {year} {2010})}\BibitemShut {NoStop}%
\bibitem [{\citenamefont {White}\ \emph {et~al.}(2007)\citenamefont {White},
  \citenamefont {Li}, \citenamefont {Gunlycke},\ and\ \citenamefont
  {Mintmire}}]{White2007}%
  \BibitemOpen
  \bibfield  {author} {\bibinfo {author} {\bibfnamefont {C.~T.}\ \bibnamefont
  {White}}, \bibinfo {author} {\bibfnamefont {J.}~\bibnamefont {Li}}, \bibinfo
  {author} {\bibfnamefont {D.}~\bibnamefont {Gunlycke}}, \ and\ \bibinfo
  {author} {\bibfnamefont {J.~W.}\ \bibnamefont {Mintmire}},\ }\href {\doibase
  10.1021/nl0627745} {\bibfield  {journal} {\bibinfo  {journal} {Nano Lett.}\
  }\textbf {\bibinfo {volume} {7}},\ \bibinfo {pages} {825} (\bibinfo {year}
  {2007})}\BibitemShut {NoStop}%
\bibitem [{\citenamefont {Grassi}\ \emph {et~al.}(2013)\citenamefont {Grassi},
  \citenamefont {Gnudi}, \citenamefont {Imperiale}, \citenamefont {Gnani},
  \citenamefont {Reggiani},\ and\ \citenamefont {Baccarani}}]{Grassi2013}%
  \BibitemOpen
  \bibfield  {author} {\bibinfo {author} {\bibfnamefont {R.}~\bibnamefont
  {Grassi}}, \bibinfo {author} {\bibfnamefont {A.}~\bibnamefont {Gnudi}},
  \bibinfo {author} {\bibfnamefont {I.}~\bibnamefont {Imperiale}}, \bibinfo
  {author} {\bibfnamefont {E.}~\bibnamefont {Gnani}}, \bibinfo {author}
  {\bibfnamefont {S.}~\bibnamefont {Reggiani}}, \ and\ \bibinfo {author}
  {\bibfnamefont {G.}~\bibnamefont {Baccarani}},\ }\href
  {http://www.dx.doi.org/10.1063/1.4800900} {\bibfield  {journal} {\bibinfo
  {journal} {J. Appl. Phys.}\ }\textbf {\bibinfo {volume} {113}} (\bibinfo
  {year} {2013})}\BibitemShut {NoStop}%
\bibitem [{\citenamefont {Yoon}\ \emph {et~al.}(2011)\citenamefont {Yoon},
  \citenamefont {Nikonov},\ and\ \citenamefont {Salahuddin}}]{Yoon2011}%
  \BibitemOpen
  \bibfield  {author} {\bibinfo {author} {\bibfnamefont {Y.}~\bibnamefont
  {Yoon}}, \bibinfo {author} {\bibfnamefont {D.~E.}\ \bibnamefont {Nikonov}}, \
  and\ \bibinfo {author} {\bibfnamefont {S.}~\bibnamefont {Salahuddin}},\
  }\href {\doibase 10.1063/1.3589365} {\bibfield  {journal} {\bibinfo
  {journal} {Appl. Phys. Lett.}\ }\textbf {\bibinfo {volume} {98}},\ \bibinfo
  {pages} {1} (\bibinfo {year} {2011})}\BibitemShut {NoStop}%
\bibitem [{\citenamefont {Koswatta}\ \emph {et~al.}(2007)\citenamefont
  {Koswatta}, \citenamefont {Hasan}, \citenamefont {Lundstrom}, \citenamefont
  {Anantram},\ and\ \citenamefont {Nikonov}}]{Koswatta2007a}%
  \BibitemOpen
  \bibfield  {author} {\bibinfo {author} {\bibfnamefont {S.~O.}\ \bibnamefont
  {Koswatta}}, \bibinfo {author} {\bibfnamefont {S.}~\bibnamefont {Hasan}},
  \bibinfo {author} {\bibfnamefont {M.~S.}\ \bibnamefont {Lundstrom}}, \bibinfo
  {author} {\bibfnamefont {M.~P.}\ \bibnamefont {Anantram}}, \ and\ \bibinfo
  {author} {\bibfnamefont {D.~E.}\ \bibnamefont {Nikonov}},\ }\href {\doibase
  10.1109/TED.2007.902900} {\bibfield  {journal} {\bibinfo  {journal} {IEEE
  Trans. Electron Devices}\ }\textbf {\bibinfo {volume} {54}},\ \bibinfo
  {pages} {2339} (\bibinfo {year} {2007})}\BibitemShut {NoStop}%
\bibitem [{\citenamefont {Koswatta}\ \emph {et~al.}(2008)\citenamefont
  {Koswatta}, \citenamefont {Lundstrom},\ and\ \citenamefont
  {Nikonov}}]{KLN2008}%
  \BibitemOpen
  \bibfield  {author} {\bibinfo {author} {\bibfnamefont {S.~O.}\ \bibnamefont
  {Koswatta}}, \bibinfo {author} {\bibfnamefont {M.~S.}\ \bibnamefont
  {Lundstrom}}, \ and\ \bibinfo {author} {\bibfnamefont {D.~E.}\ \bibnamefont
  {Nikonov}},\ }\href {\doibase 10.1063/1.2839375} {\bibfield  {journal}
  {\bibinfo  {journal} {Appl. Phys. Lett.}\ }\textbf {\bibinfo {volume} {92}},\
  \bibinfo {pages} {1} (\bibinfo {year} {2008})}\BibitemShut {NoStop}%
\end{thebibliography}
\end{document}